\documentclass[11pt,a4paper]{article}

\usepackage{jheppub} 

\usepackage[T1]{fontenc} 
\usepackage{multirow}
\usepackage{amsmath}
\usepackage{pifont}
\usepackage{float}
\usepackage{cancel}

\allowdisplaybreaks

\usepackage{pdfsync}
\usepackage{multirow}
\usepackage{simplewick}
\usepackage{textcomp}

\usepackage{lineno} 

\usepackage{hyperref} 
\hypersetup{
    colorlinks,
    citecolor=black,
    filecolor=black,
    linkcolor=black,
    urlcolor=black
}

\RequirePackage{flushend}
\RequirePackage[numbers,sort&compress]{natbib}

\usepackage{marvosym} 

\usepackage{slashed}

\newcommand{\dhalf}{\frac{d}{2}}

\newcommand{\Ds}{\displaystyle}

\newcommand{\nn}{\nonumber}

\newcommand{\Li}{\text{Li}}

\renewcommand{\(}{\left(}
\renewcommand{\)}{\right)}
\renewcommand{\[}{\left[}
\renewcommand{\]}{\right]}

\newcommand{\OO}{\mathbb{O}}

\newcommand{\he}{{\Delta q}}

\def\II{\hbox{{1}\kern-.25em\hbox{l}}}

\title{QCD factorization for twist-three axial-vector parton quasidistributions}

\author[a]{Vladimir~M.~Braun,}
\author[b]{Yao Ji,}
\author[a]{and Alexey~Vladimirov}

\subheader{
\begin{flushright}
{\small
{SI-HEP-2021-07}
}
\end{flushright}
}

\affiliation[a]{Institut f\"ur Theoretische Physik, Universit\"at Regensburg, D-93040 Regensburg, Germany}
\affiliation[b]{Theoretische Physik 1, Naturwissenschaftlich-Technische Fakult{\"a}t,
Universit{\"a}t Siegen, 57068 Siegen,\\ Germany}

\emailAdd{vladimir.braun@ur.de}
\emailAdd{yao.ji@uni-siegen.de}
\emailAdd{alexey.vladimirov@ur.de}

\abstract{
The transverse component of the axial-vector correlation function
of quark fields is a natural starting object for lattice calculations 
of twist-3 nucleon parton distribution functions. In this work we derive the corresponding 
factorization expression in terms of twist-2 and twist-3 collinear distributions to one-loop accuracy.
The results are presented both in position space, as the factorization theorem for Ioffe-time distributions,
and in momentum space, for the axial-vector quasi- and pseudodistributions.
}

\begin{document} 

\maketitle 

\section{Introduction}

Twist-three effects originate from the quantum-mechanical interference between a single parton and a gluon-parton pair. Their investigation is exceptionally important for the understanding of the nature 
of strong interactions, but it is also a very challenging task. 
The main complication is that twist-three contributions are usually of subleading power in the hard scale, and
either contaminated by the leading-twist contributions or statistically suppressed. 
Additionally, a typical twist-three observable is sensitive only to a certain projection of 
the underlying quark-antiquark-gluon correlation function and a combination of many observables is 
required to unravel its structure. 

The structure function $g_2(x,Q^2)$ of the polarized deep-inelastic lepton-nucleon scattering (DIS) provided,   
historically, a paradigm case and the main motivation for studies of twist-three effects. In this example, the QCD description
of twist-three effects in the framework of collinear factorization or, equivalently, the operator product expansion (OPE)
was understood and worked out in some detail, see, e.g., 
\cite{Burkhardt:1970ti,Wandzura:1977qf,Kodaira:1978sh,Shuryak:1981pi,Bukhvostov:1984rns,Ratcliffe:1985mp,Balitsky:1987bk,Jaffe:1990qh,Ji:1990br,Ali:1991em,Kodaira:1994ge,Kodaira:1996md,Braun:1998id,Belitsky:1999bf,Derkachov:1999ze,Braun:2000av,Braun:2000yi,Braun:2001qx,Bluemlein:2002be}.
The experimental measurement of $g_2(x,Q^2)$ is, however, very difficult. 
The existing data are scattered~\cite{Anthony:1999py,Anthony:2002hy,Slifer:2008xu,Airapetian:2011wu,Flay:2016wie,Armstrong:2018xgk}, and the twist-three part of $g_2(x,Q^2)$ can be extracted with large uncertainties only~\cite{Sato:2016tuz}.
The first nontrivial moment of $g_2(x,Q^2)$ that is related to 
the matrix element of the local chiral-even twist-three operator of the lowest dimension, was estimated using 
QCD sum rules~\cite{Balitsky:1989jb,Stein:1994zk}, models~\cite{Balla:1997hf,Braun:2011aw}, and in lattice QCD~\cite{Gockeler:2005vw}.
A comparison of the predictions with the available data can be found in Ref.~\cite{Armstrong:2018xgk}.
In the future, high-precision measurements of $g_2(x,Q^2)$ are planned for the JLAB 12~GeV 
upgrade~\cite{Dudek:2012vr} and later at the Electron-Ion Collider (EIC)~\cite{AbdulKhalek:2021gbh}.  
 
More recently, the interest in twist-three effects was fueled by studies of transverse momentum dependent (TMD) parton distributions. 
In this case, the twist-three distributions arise in the collinear limit for many TMD distributions already at 
the leading power \cite{Boer:2003cm,Kanazawa:2015ajw,Scimemi:2018mmi,Moos:2020wvd}. 
In particular, the single transverse spin asymmetry is sensitive to the Sivers function \cite{Sivers:1989cc}, 
which matches the Qiu-Sterman function \cite{Qiu:1991pp} in the collinear limit. 
The latter is a certain integral of the same quark-antiquark-gluon correlation function that contributes 
to $g_2(x,Q^2)$, see, e.g., \cite{Kang:2011mr,Braun:2009mi,Scimemi:2019gge}. 
In this way, one can extract twist-three distributions from the transverse momentum dependence, which 
has been done recently for the Qiu-Sterman function \cite{Bury:2020vhj,Bury:2021sue}.

Lattice QCD simulations have the potential to explore the plethora of twist-three distributions more directly by the calculation of 
Euclidean correlation functions that are specially tailored to the twist-three effects. 
The lattice calculations of the leading-twist parton distributions (PDFs) 
have attracted a lot of attention, see~\cite{Lin:2020rut,Ji:2020ect} for a review. In this case, a convenient object is an equal-time 
correlation function of two quark (or gluon) fields connected by the straight Wilson line, 
which can be factorized~\cite{Izubuchi:2018srq} in terms of the corresponding PDF convoluted with a perturbatively calculable coefficient function.
The lattice ``data'' can be analyzed either directly in position space in terms of Ioffe-time distributions~\cite{DelDebbio:2020rgv}, 
or Fourier-transformed to momentum space where they are referred to as quasidistributions (qPDFs)~\cite{Ji:2013dva} 
or pseudodistributions (pPDFs)~\cite{Radyushkin:2017cyf}.
   
In Refs.~\cite{Bhattacharya:2020cen,Bhattacharya:2020xlt} the same approach was suggested to extract 
the twist-three PDF $g_T(x)$~\cite{Jaffe:1996zw} from a lattice calculation of the correlation function
\begin{align}
\label{q-object}
\langle N(p)|\bar q(z)\gamma^\mu \gamma^5 [z,0] q(0)|N(p)\rangle\,, 
\end{align}
where $z^\mu$ is a space-like vector and $[z,0]$ is the Wilson line. We refer to this object as a ``quasidistribution'' in a broad sense,
to distinguish from the parton distributions defined for the light-like separation $z^2=0$.
In this work, we derive the factorization formula for the correlation function in \eqref{q-object} to twist-three accuracy 
at next-to-leading perturbative order (NLO). 
The results are presented both in position space, as a factorization theorem for Ioffe-time distributions,
and in momentum space, for the axial-vector quasi- and pseudodistributions.

Our main results can be summarized as follows:
\begin{itemize}
\item The factorization theorem to twist-three accuracy has a more complicated structure as compared to the leading twist~\cite{Izubuchi:2018srq}.
In particular the quasidistribution \eqref{q-object} cannot be factorized in terms of the parton distribution $g_T(x)$~\cite{Jaffe:1996zw},
as conjectured in~\cite{Bhattacharya:2020cen,Bhattacharya:2020xlt}.
\item We have checked that the factorization scale dependence of the coefficient functions in our expressions agrees with
the known results on the renormalization of twist-three operators, 
cf.~\cite{Bukhvostov:1984rns,Balitsky:1987bk,Kodaira:1996md,Braun:2009mi, Ji:2014eta}.
\item The simple LO relation between the twist-two contributions to the quasidistributions~\eqref{q-object} 
with longitudinal and transverse projections of the vector index is violated at NLO. Hence the Wandzura-Wilczek relation is violated
for quasidistributions. This is different from DIS, in which case the Wandzura-Wilczek relation for twist-two contributions holds on the level of structure functions.   
\item In the limit of large number of colors, the logarithmic terms in the coefficient functions are simplified \cite{Ali:1991em,Braun:2001qx} and can be combined to express the result in terms of the quark-antiquark transverse spin distribution~$g_T(x)$. 
However, this simplification does not occur for finite terms. 
\item We consider a short-distance expansion of \eqref{q-object} which may provide a method to calculate the matrix element of the twist-three operator of the lowest dimension avoiding a complicated procedure of nonperturbative renormalization in the presence of power divergences, cf.~\cite{Gockeler:2005vw}. 
\end{itemize}

The presentation is organized as follows. Sect.~2 and Sect.~3 are introductory and contain main definitions and notations.
In Sect.~4 we formulate the factorization theorem and derive the NLO expressions at the operator level and for the position-space (Ioffe-time) distributions. In Sect.~5 the corresponding results are presented for the qPDFs and pPDfs.
The final Sect.~6 contains a discussion and outlook. Technical details are delegated to the Appendices.

\section{Definitions}

In the present work we study the twist expansion of a product of quark and antiquark fields 
\begin{eqnarray}\label{def:op-main}
\mathcal{O}^{\gamma^\mu \gamma^5}(z,0)=\text{T}\{\bar q(z)\gamma^\mu \gamma^5 [z,0]q(0)\},
\end{eqnarray}
where $z^\mu$ is a four-vector, $q$ are quark fields, and $[z,0]$ is the straight Wilson line
in the fundamental representation of the gauge group
\begin{eqnarray}
[z,0]=P\exp\Big(ig\int_{0}^{1}\!d\sigma\, z^\mu A_\mu(\sigma z)\Big).
\end{eqnarray}
For space-like separations $z^2<0$ which are relevant for lattice calculations, the time-ordering in Eq.~\eqref{def:op-main} is redundant. The Dirac matrix $\gamma^5$ is defined (in $d=4$ dimensions) as
\begin{eqnarray}\label{def:gamma5}
\gamma^5=i\gamma^0\gamma^1\gamma^2\gamma^3=\frac{-i}{4!}\epsilon^{\mu\nu\rho\sigma}\gamma_\mu\gamma_\nu\gamma_\rho\gamma_\sigma,
\end{eqnarray}
where $\epsilon$ is the Levi-Civita tensor with $\epsilon^{0123}=-\epsilon_{0123}=1$. 
Flavor indices of the quark fields are omitted for brevity.

If not stated otherwise, we tacitly assume that the nonlocal operator \eqref{def:op-main} is renormalized,
\begin{eqnarray}\label{def:ZZO}
\mathcal{O}^{\gamma^\mu \gamma^5}(z,0)&=& Z_{\mathcal O} {\cal O}_{\text{bare}}^{\gamma^\mu \gamma^5}(z,0)\,.
\end{eqnarray}
The renormalization constant $Z_{\mathcal O}$ in the $\overline{\text{MS}}$ scheme 
is known to three-loop accuracy~\cite{Chetyrkin:2003vi,Braun:2020ymy}.
It is the same for all Dirac structures. 
Both sides of Eq.~\eqref{def:ZZO} depend on the renormalization scale $\mu_R$. This dependence is not shown explicitly 
in what follows, unless it is important for understanding. 

In  renormalization  schemes  with  an  explicit  regularization  scale,  the Wilson  line  in  Eq.~\eqref{def:op-main}  
suffers  from  an  additional  linear  ultraviolet  divergence \cite{Dotsenko:1979wb} which  has  to  be  removed.   
This  can  be  done  by the renormalization of a residual mass term,  similarly to the heavy quark effective theory, 
or, alternatively, by forming a suitable  ratio of matrix elements involving the same operator \cite{Orginos:2017kos,Braun:2018brg}.
This issue is well known and does not require an extra elaboration.
 
In this work, we study only the forward matrix elements. Thus, the global positioning of the operator is unimportant, and only the distance $z$ between the fields plays a role. Without loss of generality, we can put the quark position at the origin.
The nucleon matrix element of the operator in Eq.~\eqref{def:op-main} can be parametrized in terms of three invariant functions,
\begin{align}\label{def:ITD}
\langle p,s|\mathcal{O}^{\gamma^\mu \gamma^5}(z,0)|p,s\rangle=2\Big(p^\mu \lambda_z \mathcal{G}_1(p\cdot z,z^2)&+s_T^\mu M \mathcal{G}_T(p\cdot z,z^2)
\\\nn &
+\frac{z^\mu(p\cdot z)-p^\mu z^2}{(p\cdot z)^2}\lambda_z M^2\mathcal{G}_3(p\cdot z,z^2)\Big).
\end{align}
Here, $M^2=p^2$ is the nucleon mass, $s^\mu$ is the spin vector, $(s\cdot p)=0$, $s^2=-1$, and
\begin{eqnarray}
\lambda_z= M\frac{(s\cdot z)}{(p\cdot z)}.
\end{eqnarray}
We refer the functions $\mathcal{G}$ as Ioffe-time quasidistributions (qITDs) \cite{Ioffe:1969kf,Braun:1994jq,Orginos:2017kos}, and the variable
\begin{align}
\zeta =  p_z = (p\cdot z),
\end{align}
as the Ioffe time~\cite{Ioffe:1969kf}.

The subscript $T$ indicates the projection onto the transverse plane, which is orthogonal to  $z^\mu$ and $p^\mu$:
\begin{align}
s_T^\mu = g_T^{\mu\nu} s_\nu\,,
\end{align} 
with
\begin{eqnarray}\label{def:gTtensor}
g_T^{\mu\nu}=g^{\mu\nu}-(z^\mu p^\nu+p^\mu z^\nu)\frac{p_z}{p_z^2-z^2 p^2}+p^\mu p^\nu \frac{z^2}{p_z^2-z^2p^2}+z^\mu z^\nu \frac{p^2}{p_z^2-z^2 p^2}.
\end{eqnarray}
The transverse projection of the metric tensor in four dimensions satisfies
\begin{eqnarray}
p_\mu g_T^{\mu\nu}=z_\mu g_T^{\mu\nu}=0\,,\qquad g_T^{\mu\nu}g_{T\nu}^{~~\rho}=g_T^{\mu\rho},\qquad g^{\mu}_{T\mu}=2.
\end{eqnarray}
For further use we also introduce the anti-symmetric tensor
\begin{eqnarray}\label{def:epsilonT}
\epsilon^{\mu\nu}_T=\frac{p_\alpha z_\beta \epsilon^{\alpha\beta \mu\nu}}{\sqrt{p_z^2-z^2p^2}}\,,
\end{eqnarray}
such that 
\begin{eqnarray}\label{th:epsilonT-properties}
\epsilon_T^{\mu\nu}\epsilon_{T\nu}^{~~~\rho}=-g_T^{\mu\rho},\qquad \epsilon^{\mu\nu}_T\epsilon_{T\mu\nu}=2\,.
\end{eqnarray}
If the four-vectors $z^\mu$ and $p^\mu$ are positioned in the $(t,z)$-plane, 
then $\epsilon^{12}_T=-\epsilon_T^{21}=1$ (and other components vanish). 
Let us note that the terms $\sim z^2p^2$ can be omitted, since in the following we study the limit $z^2 p^2/p_z^2\to 0$. However, we keep these terms in (\ref{def:gTtensor}) and (\ref{def:epsilonT}) to ensure exact orthogonality conditions.

The three qITDs $\mathcal{G}_1$, $\mathcal{G}_T$ and $\mathcal{G}_3$ at small $z^2$ can be matched to collinear PDFs. 
This matching is studied in great detail for $\mathcal{G}_1$, where only the twist-two collinear PDF
contributes at accuracy $\mathcal{O}(z^2)$. The corresponding coefficient function is known to two 
loops (NNLO)~\cite{Chen:2020arf,Chen:2020ody,Li:2020xml}.
The factorized expressions for $\mathcal{G}_T$ and $\mathcal{G}_3$ are more complicated. At the same power accuracy, they
receive contributions from collinear PDFs of twists 2,3 and twists 2,3,4, respectively. 
In this paper we derive the leading-power factorization theorem 
for the qITD  $\mathcal{G}_T$ distribution to NLO (one-loop) accuracy. 
It incorporates twist-2 and twist-3 collinear distributions. 
Starting from this result, the coefficient functions for qPDFs and pPDFs can be obtained by the appropriate 
Fourier transform. The necessary definitions are given in the corresponding sections.  

\section{Light-ray operators}

Operator product expansion (OPE) can be conveniently organized in terms of the generating functions
of (renormalized) local operators~\cite{Anikin:1978tj,Anikin:1979kq,Balitsky:1987bk,Mueller:1998fv,Balitsky:1990ck,Geyer:1999uq}  
\begin{align}
\label{OO}
 [{\OO}^{\gamma^\mu \gamma^5}(z,0)]^{\mu_F} =  [\bar q(z)\gamma^\mu \gamma^5 [z,0] q(0)]^{\mu_F} = 
\sum_{k=0}^\infty\frac{1}{k!} z^{\mu_1} \ldots z^{\mu_k} [{O}^{\gamma^\mu \gamma^5}_{\mu_1\ldots\mu_k}(0)]^{\mu_F},
\end{align}
where
\begin{align}
 {O}^{\gamma^\mu \gamma^5}_{\mu_1\ldots\mu_k}(0) = \bar q (0)\stackrel{\leftarrow}{D}_{\mu_1} \ldots \stackrel{\leftarrow}{D}_{\mu_k}
\gamma^\mu \gamma^5 q(0)\,.  
\end{align}
The superscript $\mu_F$ indicates the renormalization scale for the operator. In what follows, 
we often suppress the scale dependence not to overload the notation. 
Importantly, the operator ${\OO}^{\gamma^\mu \gamma^5}(z,0)$ in Eq.~\eqref{OO} and $\mathcal{O}^{\gamma^\mu \gamma^5}(z,0)$ in Eq.~\eqref{def:op-main}
are different beyond the tree level.
 
Local operators on the r.h.s. of Eq.~\eqref{OO} can be decomposed into a sum of contributions with different 
geometric twist (dimension minus spin). In particular, the twist-two operators are obtained by symmetrization over all
Lorentz indices and subtraction of traces. We define the twist-two projection of the nonlocal operator \eqref{OO} as the
generating function for (renormalized) local twist-two operators
\begin{align}
\label{OOlt}
 [{\OO}^{\gamma^\mu \gamma^5}(z,0)]_{\rm tw2}&= \sum_{k=0}^\infty\frac{1}{k!} z^{\mu_1} \ldots z^{\mu_k} [{O}^{\gamma^\mu \gamma^5}_{\mu_1\ldots\mu_k}(0)]_{\rm tw2}
\notag\\&= 
\sum_{k=0}^\infty\frac{1}{k!(k+1)!}
\frac{\partial}{\partial n^\mu} \Big(z_\rho \frac{\partial}{\partial n^\rho}\Big)^k {O}^{\slashed{n} \gamma^5}_{n\ldots n}(0)
\notag\\&= 
\sum_{k=0}^\infty\frac{1}{(k+1)!}
\frac{\partial}{\partial n^\mu} \Big(z_\rho \frac{\partial}{\partial n^\rho}\Big)^k 
{\OO}^{\slashed{n} \gamma^5}(n,0)\Big|_{n=0}, 
\end{align}    
where $n^\mu$ is an auxiliary light-like vector, $n^2=0$, and 
${O}^{\slashed{n} \gamma^5}_{n\ldots n}(0) = n^{\mu_1}\ldots  n^{\mu_k}  {O}^{\slashed{n} \gamma^5}_{\mu_1\ldots \mu_k}(0)$.
Note that operator ${\OO}^{\slashed{n} \gamma^5}(n,0)$ in the last line is a nonlocal operator defined as in Eq.~\eqref{OO}
but at the light-like separation $z^\mu\to n^\mu$ between the fields --- the light-ray operator. 
The twist-two projection \eqref{OOlt} of the nonlocal operator \eqref{OO} satisfies the equations \cite{Balitsky:1990ck}
\begin{align}
 \frac{\partial}{\partial z^\mu}  [{\OO}^{\gamma^\mu \gamma^5}(z,0)]_{\rm tw2} = 0\,,
\qquad 
 \frac{\partial}{\partial z_\rho}\frac{\partial}{\partial z^\rho}[{\OO}^{\gamma^\mu \gamma^5}(z,0)]_{\rm tw2} =0\,,    
\end{align} 
and can be written more explicitly in several different representations 
\cite{Balitsky:1987bk,Mueller:1998fv,Balitsky:1990ck,Geyer:1999uq,Braun:2011dg}.
Similarly, it is possible to construct the projectors onto twist-three and higher-twists operators.

In this work we consider the transverse component ${\OO}^{\gamma_T^\mu \gamma^5}(z,0) =   g^{\mu\nu}_T {\OO}^{\gamma_\nu \gamma^5}(z,0)$ and neglect twist-four contributions. 
Hence, we can omit terms $\mathcal{O}(z^2)$ corresponding to the subtraction of traces. To this accuracy it is sufficient to use the following decomposition
\begin{align}
\label{t23}
  {\OO}^{\gamma^\mu \gamma^5}(z,0) =  [{\OO}^{\gamma^\mu \gamma^5}(z,0)]_{\rm tw2} + [{\OO}^{\gamma^\mu \gamma^5}(z,0)]_{\rm tw3} \,,
\end{align}
where~\cite{Balitsky:1990ck}
\begin{eqnarray}
\label{t=2}
[{\OO}^{\gamma^\mu \gamma^5}(z,0)]_{\rm tw2} & =& \int_0^1d\alpha \frac{\partial}{\partial z_\mu} {\OO}^{\slashed{z} \gamma^5}(\alpha z,0) \, ,  
\\
\label{t=3}
[{\OO}^{\gamma^\mu \gamma^5}(z,0)]_{\rm tw3} &=&  \int_0^1 d\alpha \, z^\rho 
\Big(\frac{\partial}{\partial z^\rho} {\OO}^{\gamma^\mu \gamma^5}(\alpha z ,0)-\frac{\partial}{\partial z^\mu} {\OO}^{\gamma^\rho \gamma^5}(\alpha z,0) 
\Big)
\\\nn
&=&
-\frac{1}{2}\int_0^1 du \int_{0}^u dv 
\(
(u-v)\mathbb{T}^\mu_-(uz,vz,0) - v\mathbb{T}^\mu_+(uz,vz,0)  
\).
\end{eqnarray}
Here and below we use the notation
\begin{eqnarray}\label{def:qGq-operators+}
\mathbb{T}^\mu_+(az,bz,cz)&=&
i g\, \bar q(az)\,\gamma^\rho \slashed{z}\gamma^\mu \gamma^5 \, F_{\rho z}(bz)q(cz)
\\\nn &=&-i g\, g^{\mu\rho}_T \bar q(az) \slashed{z}\(F_{\rho z }(bz)\gamma_5 + i\widetilde F_{\rho z }(bz)\)q(cz),
\\\label{def:qGq-operators-}
\mathbb{T}^\mu_-(az,bz,cz)&=&
i g\, \bar q(az)\,\gamma^\mu \slashed{z}\gamma^\rho \gamma^5 \, F_{\rho z}(bz)q(cz)
\\\nn &=&-i g\, g^{\mu\rho}_T \bar q(az) \slashed{z}\(F_{\rho z }(bz)\gamma_5 - i\widetilde F_{\rho z }(bz)\)q(cz),
\end{eqnarray}
where $g$ is the QCD coupling constant, $F_{\mu \nu}$ is the gluon field-strength tensor, $F_{\mu z}=F_{\mu \nu}z^\nu$, and $\widetilde F_{\mu \nu} = \frac{1}{2}\epsilon_{\mu\nu\alpha\beta} F^{\alpha\beta}$. The Wilson lines connecting the three fields in the quark-antiquark-gluon operators \eqref{def:qGq-operators+}, \eqref{def:qGq-operators-}
are implied. Let us emphasize that Eqs.~\eqref{t=2} and \eqref{t=3} are exact operator identities for renormalized operators. We also note that the $\mathbb{T}^\mu_\pm$ operators involve the QCD coupling, hence $[{\OO}^{\gamma^\mu \gamma^5}(z,0)]_{\rm tw3}$ vanishes in the free theory. So, the twist-three effects arise due to quark-gluon interactions.

Nucleon matrix elements of the nonlocal operators (\ref{OO}, \ref{def:qGq-operators+}, \ref{def:qGq-operators-}) define parton distributions.
Neglecting terms  $\mathcal{O}(z^2)$
\begin{align}
\label{def:g1}
\langle p,s|\OO^{\slashed{z} \gamma^5}(z,0)|p,s\rangle& = 2 \lambda_z (p\cdot z) \int_{-1}^1 \!dx\,  e^{ix (p\cdot z) } \he(x)
\equiv   2\lambda_z\, \zeta\, \widehat \he (\zeta), 
\\
\label{def:gT}
\langle p,s|\OO^{\gamma^\mu_T \gamma^5}(z,0)|p,s\rangle &= 2 s^\mu_TM \int_{-1}^1 \! dx\, e^{ix(p\cdot z)} g_T(x)
\equiv 2 s^\mu_TM\, \widehat g_T(\zeta).
\end{align}
Here and in what follows, we use the ``hat'' notation for the PDFs in position space, dubbed Ioffe-time distributions~\cite{Ioffe:1969kf,Braun:1994jq}. 
The PDF $\he(x)$ for $x>0$ and $x<0$ defines helicity distribution of quarks and antiquarks (with the sign minus) in  the nucleon,
respectively. The function $g_T(x)$ can be decomposed into twist-two and twist-three contributions \eqref{t23}. 
Using \eqref{t=2} one obtains the twist-two part
\begin{align}
\widehat{g}^{\text{tw2}}_T(\zeta)&=\int_0^1 d\alpha\, \widehat\he(\alpha \zeta), 
\notag\\
{g}^{\text{tw2}}_T(x) &= 
\theta(x) \int_x^1\frac{dy}{y} \he(y)
- \theta(-x) \int_{-1}^x\frac{dy}{y} \he(y)\,.
\label{gT-tw2}
\end{align}
This is the celebrated Wandzura-Wilczek relation~\cite{Wandzura:1977qf}.

For the twist-three part using Eq.~\eqref{t=3} we get
\begin{align}
\widehat{g}^{\text{tw3}}_T(\zeta)&=
\zeta^2\int_0^1 d\alpha\int_\alpha^1 d\beta \bar \beta \Big(\widehat{S}^-(\zeta,\beta \zeta,\alpha\zeta)+\widehat S^+(\bar \alpha \zeta,\bar \beta \zeta,0)\Big),
\label{gT-tw3}
\end{align}
where $\bar\alpha =1-\alpha$, etc., and $\widehat{S}^\pm$ are the twist-three quark-antiquark-gluon correlation functions (in position space) defined as 
\begin{eqnarray}\nn\label{def:S+}
\langle p,s| \mathbb{T}^\mu_+ (az,bz,cz)|p,s\rangle &=& 4 s^\mu_T\zeta^ 2M \,\widehat{S}^+(a\zeta,b\zeta,c\zeta),
\\\nn\label{def:S-}
\langle p,s|\mathbb{T}^\mu_- (az,bz,cz)|p,s\rangle &=& - 4 s^\mu_T \zeta^ 2M \,\widehat{S}^-(a\zeta,b\zeta,c\zeta).
\end{eqnarray}

The position-space PDFs $\widehat{S}^\pm$ are related to the PDFs in  momentum fraction space by the Fourier transformation 
\begin{align}\label{def:twist-3-Fourier}
\widehat{S}^\pm(\zeta_1,\zeta_2,\zeta_3)& =\int [dx]\, e^{-i(\zeta_1x_1+\zeta_2x_2+\zeta_3x_3)}S^\pm(x_1,x_2,x_3),
\end{align}
where 
\begin{align}\label{def:[dx]}
\int [dx]&=\int_{-1}^1 \!dx_1dx_2dx_3\,\delta(x_1+x_2+x_3)\,.
\end{align}
The functions $S^\pm$ obey the symmetry relation~\cite{Kodaira:1994ge,Braun:2001qx}
\begin{align}\label{def:S-sym}
S^\pm(x_1,x_2,x_3)&= S^\mp(-x_3,-x_2,-x_1),
\end{align}
and similar for $\widehat{S}^\pm$. 
It follows that $\widehat{S}^-(\zeta,\beta \zeta,\alpha\zeta)=\widehat{S}^+(\bar \alpha \zeta,\bar \beta \zeta,0)$, and, therefore, 
the two terms in \eqref{gT-tw3} are equal to each other. Thus we write simply 
\begin{align}
\widehat{g}^{\text{tw3}}_T(\zeta)&=
2\zeta^2\int_0^1 d\alpha\int_\alpha^1 d\beta\, \bar \beta\, \widehat{S}^-(\zeta,\beta \zeta,\alpha\zeta)\,,
\label{gT-tw3-simpl}
\end{align}
and, going over to the momentum fraction representation, obtain
\begin{align}
\label{gT-tw3-x}
g^{\text{tw3}}_T(x) &=
2\int [dx]\int_0^1 \! d\alpha \(\frac{\delta(x+\alpha x_1)}{x_1x_3}+\frac{\delta(x+x_1+\alpha x_2)}{x_2x_3}+\frac{\delta(x+x_1)}{x_1x_2}\)S^-(x_1,x_2,x_3).
\end{align}
In this expression one can get rid of two integrations using the delta-functions, but the resulting expression is unwieldy due to a multitude of integration regions.
For comparison, the corresponding expression in position space \eqref{gT-tw3-simpl} is rather compact. This situation is rather general.
For this reason, we carry out the major part of the calculations in position space, and go over to momentum fractions only in the very end.

The relation $x_1+x_2+x_3=0$ (\ref{def:[dx]}) implies that the twist-three PDFs are functions of two variables. 
However, the three-variable notation used here is more convenient for many reasons. First, it simplifies the symmetry relations (\ref{def:S-sym}).  
Second, twist-three correlation functions have different parton interpretation in each kinematic 
domain $x_i\lessgtr 0$~\cite{Jaffe:1983hp} and can most naturally be presented using three-component
barycentric coordinates, see~\cite{Braun:2009mi,Scimemi:2019gge}.

There is no single established notation for twist-three PDFs. Our definition is common in the literature on the DIS
structure function $g_2(x,Q^2)$, e.g.,~\cite{Shuryak:1981pi,Mueller:1997yk,Braun:2001qx}. 
In applications to semi-inclusive reactions~\cite{Braun:2009mi,Scimemi:2019gge,Ji:2006vf,Kang:2011mr,Scimemi:2018mmi,Moos:2020wvd},
it is customary to use a different pair of twist-three PDFs  defined as
\begin{eqnarray}
\langle p,s|\bar q(az)\,\slashed{z}\,F_{\mu z}(bz)q(cz)|p,s\rangle &=&2\epsilon^{\mu\nu}_T s_\nu \zeta^2 M\,\widehat{T}(a\zeta,b\zeta,c\zeta)\,,
\\
\langle p,s|\bar q(az)\,\slashed{z}\gamma^5 \,F_{\mu z}(bz)q(cz)|p,s\rangle &=&-2is_T^\mu\zeta^2M\,\Delta\widehat{T}(a\zeta,b\zeta,c\zeta)\,.
\end{eqnarray}
The corresponding momentum fraction distributions $T$ and $\Delta T$ (defined as in \eqref{def:twist-3-Fourier}) are related to 
$S^\pm$ as
\begin{eqnarray}
S^\pm(x_1,x_2,x_3)&=&\frac{1}{2}\(-T(x_1,x_2,x_3)\pm\Delta T(x_1,x_2,x_3)\).
\end{eqnarray}
The $T$ and $\Delta T$ functions satisfy the following symmetry relations 
\begin{align}
T(x_1,x_2,x_3)=T(-x_3,-x_2,-x_1)\,,\qquad \Delta T(x_1,x_2,x_3)=-\Delta T(-x_3,-x_2,-x_1)\,.
\end{align}
A detailed comparison between different notations can be found in Refs.~\cite{Braun:2009mi,Scimemi:2019gge}. 

The correlation functions $S^\pm$ (or, equivalently, $T$ and $\Delta T$) are scale-dependent.
Their evolution is autonomous, in the sense that it does not involve other distributions 
(apart from the gluon twist-3 distribution in the singlet case, which we do not discuss in this work). 
The evolution equation takes the form
\begin{eqnarray}\label{th:evolution-gen}
\frac{d S^\pm(x_1,x_2,x_3;\mu_F)}{d\ln \mu_F^2}=\int [dy]\, K^\pm(x,y;\alpha_s(\mu_F))\,S^\pm(y_1,y_2,y_3;\mu_F),
\end{eqnarray}
where $K$ is a kernel which, in general, depends on six variables $x= \{x_1,x_2,x_3\}$ and $y = \{y_1,y_2,y_3\}$. 
The leading-order (LO) expression for the kernel $K$ can be found in \cite{Braun:2009mi,Ji:2014eta}. 
This expression is, again, relatively compact in position space but rather lengthy in terms of the momentum fractions.
For completeness we present the position-space kernel~\cite{Braun:2009mi} in App.~\ref{app:evolution_kernel}.
On the contrary, the evolution of the function $g_T$ is not autonomous, since the projection of variables ($x_1$,$x_2$,$x_3$) onto a 
single momentum fraction $x$ is not an eigentransformation of the evolution kernel (\ref{th:evolution-gen}), except 
for the large-$N_c$ regime (and in LO only), where the function $g_T$ obeys the DGLAP-type evolution equation~\cite{Ali:1991em,Braun:2001qx}.

\section{QCD factorization for quasidistributions}
At tree level ${\OO}^{\gamma^\mu \gamma^5}(z,0) = \mathcal{O}^{\gamma^\mu \gamma^5}(z,0)$ and therefore
\begin{align}
\label{g1-tree}
\mathcal{G}_1(\zeta,z^2)&=\widehat\he(\zeta;\mu_F)\, +\ldots~,
\\
\mathcal{G}_T(\zeta,z^2)&=\widehat{g}_T(\zeta;\mu_F)\, + \ldots
\notag\\&= \int_0^1 d\alpha\, \widehat\he(\alpha \zeta;\mu_F) +
2\zeta^2\int_0^1 \! d\alpha\int_\alpha^1 d\beta \bar \beta\, \widehat{S}^-(\zeta,\beta \zeta,\alpha\zeta;\mu_F) +\ldots~,
\label{gT-tree}
\end{align}
where the ellipses stand for higher-twist $\mathcal{O}(z^2)$ corrections. Including higher-order perturbative corrections these
expressions can be generalized to the following factorization theorems:
\begin{align}
\mathcal{G}_1(\zeta,z^2)&=\int_0^1d\alpha\, C_1(\alpha, z^2\mu_F^2)\,\widehat\he(\alpha\zeta;\mu_F)\, +\ldots~,
\label{g1-fac}
\\
\mathcal{G}_T(\zeta,z^2)& =  \int_0^1 d\alpha\,C_T(\alpha, z^2\mu_F^2)\, \widehat\he(\alpha \zeta;\mu_F)
\\\nn &\qquad
 +
2\zeta^2\int_0^1 d\alpha\int_0^1 d\beta\,
C_-(\alpha, \beta,  z^2\mu_F^2)\widehat{S}^-(\zeta,\beta \zeta,\alpha\zeta;\mu_F)
 +\ldots~,
\label{gT-fac}
\end{align}
where the coefficient functions $C_k$ are given by a series expansion in the QCD coupling
\begin{align}
a_s  = \frac{\alpha_s(\mu_F)}{4\pi}\,.
\end{align}
We write
\begin{align}
 C_i = C_i^{(0)} + a_s\, C_i^{(1)} + \ldots  
\end{align}
with tree-level expressions
\begin{align}
  C_1^{(0)} = \delta(1-\alpha)\,,\qquad C_T^{(0)} = 1\,, \qquad C_{-}^{(0)} = \bar\beta\,.
\end{align}
The validity of QCD factorization for qITDs $\mathcal{G}_1$, $\mathcal{G}_T$ is a direct consequence of the existence of the operator product expansion.
The factorization theorems for quasidistributions in momentum space are obtained by the Fourier transform of above expressions and do not require any additional justification. The corresponding expressions for qPDFs and pPDFs are derived in Sec.~\ref{sec:qPDF+pPDF}.   

The main purpose of this work is the calculation of coefficient functions $C_{k}$ to one-loop accuracy.
The calculation of higher-twist contributions can be conveniently done using the 
background field technique pioneered by Schwinger \cite{Schwinger:1951xk} and adapted to the applications in nonabelian gauge theories in Refs.~\cite{Abbott:1980hw,Abbott:1981ke,Shuryak:1981kj,Novikov:1983gd,Balitsky:1987bk,Balitsky:1990ck,Balitsky:1998ya,Scimemi:2019gge}.
The basic idea is explained briefly in what follows\footnote{see also the book~\cite{Pascual:1984zb}.}. 
Explicit examples of calculations using two different versions of this approach are presented in the Appendix~\ref{app:diagrams}.

Thanks to the exact relations in \eqref{gT-tw2}, \eqref{gT-tw3}, the twist-two contribution to $\mathcal{G}_T$ 
in Eq.~\eqref{gT-fac} can be rewritten in terms of $\widehat g_T^{\mathrm{tw2}}$, 
and a part of twist-three contributions in terms of the two-particle distribution $\widehat g_T^{\mathrm{tw3}}$:
\begin{align}
\mathcal{G}_T(\zeta,z^2)& =  
\int_0^1 d\alpha\,\mathbf{C}_T(\alpha, z^2\mu_F^2)\, \widehat{g}_T^{\mathrm{tw2}}(\alpha \zeta;\mu_F)
+ 
\int_0^1 d\alpha\,\mathbf{C}_{\mathrm{2pt}}(\alpha, z^2\mu_F^2)\, \widehat{g}_T^{\mathrm{tw3}}(\alpha \zeta;\mu_F)
\notag \\ &\qquad
 +
2\zeta^2\int_0^1 d\alpha\int_0^1 d\beta\,
\mathbf{C}_{\mathrm{3pt}}(\alpha, \beta,  z^2\mu_F^2)\widehat{S}^-(\zeta,\beta \zeta,\alpha\zeta;\mu_F)\, ,
\label{gT-factor} 
\end{align}
with $ \mathbf{C}_T^{(0)} = \delta(1-\alpha)$. One should have in mind that 
the separation of the two-particle and three-particle contributions of twist three is not unique.
We discuss this possibility and its limitations in Sect.~\ref{NLOqITDs}.

\subsection{Background field technique}

The separation of coefficient functions and operator matrix elements in a certain amplitude can be understood 
in the spirit of Wilson's approach  to the renormalization group as integrating out the high-momentum 
degrees of freedom. We introduce the scale $\mu$ and define ``fast'' and ``slow'' fields as 
modes with momenta $p>\mu$ and $p<\mu$, respectively:
\begin{eqnarray}
q(x)\to q(x,\mu)+\psi(x,\mu),\qquad A_\nu(x)\to A_\nu(x,\mu)+B_\nu(x,\mu),
\end{eqnarray}
where $\psi$ and $B$ are the ``fast'', and $q$ and $A$ are the ``slow'' components.
For any gauge-invariant operator $\mathcal{O}$ 
one can integrate over the ``fast'' fields giving rise to an effective
operator that only depends on ``slow'' degrees of freedom
\begin{eqnarray}\label{th:background1}
\mathcal{O}_{\text{eff}}(q,A)=\int [D\bar \psi D\psi DB] \mathcal{O}(q+\psi,A+B)e^{iS[q+\psi,A+B]-iS[q,A]},
\end{eqnarray}
where $S$ is the QCD action. In this expression $q$ and $A$ can be considered as given fields which satisfy 
classical QCD equations of motion (EOM). The background field technique~\cite{Abbott:1980hw,Abbott:1981ke} 
is the method to evaluate such integrals paying due attention to gauge invariance.  
The presence of background field modifies the structure of the gauge-fixing term in the action. The analog of the conventional covariant gauge fixing condition for the ``fast'' fields is  \cite{Abbott:1980hw,Abbott:1981ke} 
\begin{eqnarray}
(\partial_\mu \delta^{AC}+g f^{ABC}A_\mu^B)B^{\mu,C}=D_\mu[A]B^\mu =0,
\end{eqnarray}
where $D_\mu[A]$ is the covariant derivative in the background field. 
The expression for the action $S[q+\psi,A+B]-S[q,A]$ in the background field gauge can be found 
in Ref.~\cite{Abbott:1980hw}. 

The major advantage of the background field method in our context is that the functional integral \eqref{th:background1} is invariant under the local gauge 
transformations of ``slow'' (classical) fields. This observation greatly simplifies the calculation as it implies that one can use \emph{any} suitable gauge for the background fields, and in particular a ``physical'' gauge where the gluon field can be expressed directly in terms of the strength tensor. 

There are several methods to evaluate functional integrals in background fields, which have their advantages
and drawbacks. We have performed the whole calculation using two techniques. The first approach, hereafter referred to as 
\textit{method A}, uses traditional perturbation theory for the background field action~\cite{Abbott:1980hw}
and axial gauge $z^\mu A_\mu(x)=0$ for the classical field. The second approach, \textit{method B}, uses 
the expressions for the light-cone expansion of the quark and gluon propagators in the 
background field~\cite{Balitsky:1987bk} and Fock-Schwinger gauge $x^\mu A_\mu(x)=0$.
Intermediate expressions in these two methods have a very different structure. The final results are in agreement, which
provides a strong check of their correctness.  
For pedagogical purposes we present a detailed calculation of a certain gauge-invariant subset 
of diagrams using both approaches in App.~\ref{app:diagrams}.

\subsection{Renormalization factors and treatment of $\gamma_5$}

We use the dimensional regularization with $d=4-2\epsilon$ and (modified) minimal subtraction scheme. 
Calculating the relevant Feynman diagrams in the presence of background fields we obtain the expression for the
bare qPDF operator \eqref{def:op-main} in terms of the bare light-ray operators.  The result has the following 
schematic structure
\begin{align}\label{ren:genstr1}
\mathcal{O}^{\slashed{z}\gamma^5}_{\text{bare}}(z,0)&=
C_1^{\text{bare}}\otimes [\mathbb{O}_{\text{bare}}^{\slashed{z}\gamma^5}(z)]^{\text{tw2}} + \ldots,
\notag\\
\mathcal{O}^{\gamma_T^\mu\gamma^5}_{\text{bare}}(z,0)&=
C_T^{\text{bare}}\otimes [\mathbb{O}_{\text{bare}}^{\gamma_T^\mu\gamma^5}(z)]^{\text{tw2}}
+ C_+^{\text{bare}}\otimes\mathbb{T}^\mu_{+\text{bare}}(z)
+ C_-^{\text{bare}}\otimes\mathbb{T}^\mu_{-\text{bare}}(z)+\ldots,
\end{align}
where the bare coefficient functions depend on $\epsilon$ and are singular at $\epsilon\to0$.
The $1/\epsilon$ terms in the coefficient functions are due to ultraviolet (UV) and infrared (IR) singularities and
are removed by the renormalization procedure. In the present case, it is not necessary to distinguish UV and IR poles during the calculation.

The UV singularity is removed by the (multiplicative) renormalization of the qPDF operator, Eq.~\eqref{def:ZZO}.
To one-loop accuracy~\cite{Shifman:1987rj}
\footnote{The usual $\overline{\text{MS}}$-scheme factor $(e^{\gamma_E}/(4\pi))^\epsilon$ is always implied.}
\begin{eqnarray}\label{def:ZO}
Z_{\mathcal{O}}=1-a_s\frac{3C_F}{\epsilon},
\end{eqnarray}
where $C_F=(N_c^2-1)/2N_c$ is the quadratic Casimir operator in the fundamental representation.

The IR singularities in the coefficient functions are removed by the renormalization of light-ray operators.
For twist-two~\cite{Balitsky:1987bk}
\begin{align}
 \mathbb{O}^{\slashed{z}\gamma^5}(z) &= Z_{\text{tw2}}\otimes \mathbb{O}_{\text{bare}}^{\slashed{z}\gamma^5}(z)
\notag\\&=
\mathbb{O}_{\text{bare}}^{\slashed{z}\gamma^5}(z,0)+\frac{2a_sC_F}{\epsilon}\int_0^1 d\alpha \frac{1+\alpha^2}{1-\alpha}\(\mathbb{O}_{\text{bare}}^{\slashed{z}\gamma^5}(\bar \alpha z,0)-\mathbb{O}_{\text{bare}}^{\slashed{z}\gamma^5}(z,0)\).
\end{align}
The coefficient of $1/\epsilon$ in this expression is the renown DGLAP kernel in the position-space representation. 
The renormalization factors of twist-3 operators $\mathbb{T}_\pm^\mu$ have similar structure. 
The necessary expressions can be found in \cite{Braun:2009mi}. For readers' convenience we collect them 
in App.~\ref{app:evolution_kernel}, see Eq.~\eqref{app:kernel1}. 

Inserting $\II =Z^{-1}_{\mathbb{O}}\otimes Z_{\mathbb{O}}$ in between the coefficient functions and the operators 
in each term in Eq.~\eqref{ren:genstr1}, we take the limit $\epsilon\to 0$ and obtain the final result
\begin{align}\label{ren:genstr2}
\mathcal{O}^{\slashed{z}\gamma^5}(z,0;\mu_R)&=
C_1(\mu_R,\mu_F)\otimes [\mathbb{O}^{\slashed{z}\gamma^5}(z;\mu_F)]^{\text{tw2}} + \mathcal{O}(z^2),
\notag\\
\mathcal{O}^{\gamma_T^\mu\gamma^5}(z,0;\mu_R)&=
C_T(\mu_R,\mu_F)\otimes [\mathbb{O}^{\gamma_T^\mu\gamma^5}(z;\mu_F)]^{\text{tw2}}
\notag\\&\quad
+ C_+(\mu_R,\mu_F)\otimes\mathbb{T}^\mu_{+}(z;\mu_F)
+ C_-(\mu_R,\mu_F)\otimes\mathbb{T}^\mu_{-}(z;\mu_F)+ \mathcal{O}(z^2),
\end{align}
where
\begin{eqnarray}\label{C-renormalization}
C_i(\mu_R,\mu_F)=\lim_{\epsilon\to0} Z_{\mathcal{O}}(\mu_R) C_i \otimes Z^{-1}_{i}(\mu_F)\,.
\end{eqnarray}
Here $Z_i$ stands for the renormalization kernel for the appropriate operator.
This expression involves two scales. The dependence on the renormalization scale $\mu_R$ is governed
by the anomalous dimension (\ref{def:ZO}). 
The dependence on the factorization scale $\mu_F$ is canceled between 
the coefficient functions and light-ray operators. In the following we set
\begin{align}
\mu_R=\mu_F=\mu\,.
\end{align}

The final comment concerns the treatment of $\gamma_5$. 
The usual  $\overline{\text{MS}}$ scheme is defined in such a 
way (see, e.g., \cite{Matiounine:1998re,Moch:2014sna,Gutierrez-Reyes:2017glx}) 
that the renormalization of flavor-nonsinglet  vector and axial-vector operators is assumed to be the same. 
This is achieved by starting with a suitable $\gamma_5$ definition in $d=4-2\epsilon$ 
dimensions~\cite{tHooft:1972tcz,Breitenlohner:1977hr,Larin:1993tq}
aided by a finite renormalization that effectively restores the 
anticommutation property $\{\gamma_\mu,\gamma_5\}=0$. 
This procedure is mandatory for flavor-singlet operators, but whether it has to be followed 
for flavor-nonsinglet operators depends on the method how the calculation is done.
In our case we can simply assume $\{\gamma^5,\gamma^\mu\}=0$ as $\gamma_5$  
does not appear in traces in which case using anticommutation property leads to algebraic inconsistencies.

\subsection{One-loop coefficient functions for light-ray operators}

In this section we present the results for bare coefficient functions of twist-two and twist-three operators. 
We have done the calculations using two versions of the background field technique and got the same result,
see App.~\ref{app:diagrams} for the details.
Below we write the results for the coefficient functions separating an overall factor 
\begin{align}
    C_i^{\text{bare}} = C_i^{(0)} + a_s\Gamma(-\epsilon)\(\frac{-z^2}{4}\)^\epsilon \widetilde C_i^{(1)\text{bare}}\,, \qquad i= 1,T,\pm
\end{align}
We also omit the ``bare'' superscript in order not to overload the notation.

Twist-two contributions:
\begin{align}\label{OPE:twist2}
\widetilde C_1^{(1)}\otimes [\mathbb{O}^{\slashed{z} \gamma^5}(z,0)]_{\text{tw2}}&=
2 C_F \biggl\{ \int_0^1 d\alpha 
\Big[\frac{2(1-\epsilon)}{1-2\epsilon}\frac{\alpha^{2\epsilon}-\alpha}{\alpha}\([\mathbb{O}^{\slashed{z}\gamma^5}(\bar \alpha z,0)]_{\text{tw2}}- [\mathbb{O}^{\slashed{z}\gamma^5}(z,0)]_{\text{tw2}}\)
\notag\\ &\quad
+(1-\epsilon)(1-2\epsilon)\alpha\, [\mathbb{O}^{\slashed{z}\gamma^5}(\bar \alpha z,0)]_{\text{tw2}}\Big]
-\frac{2(1-\epsilon)}{1-2\epsilon}[\mathbb{O}^{\slashed{z}\gamma^5}(z,0)]_{\text{tw2}}\biggr\},
\notag\\
\widetilde C_T^{(1)}\otimes [\mathbb{O}^{\gamma_T^\mu \gamma^5}(z,0)]_{\text{tw2}}&=
\widetilde C_1^{(1)}\otimes [\mathbb{O}^{\gamma_T^\mu \gamma^5}(z,0)]_{\text{tw2}}
+
4C_F\epsilon(1-\epsilon)\int_0^1 d\alpha \, \alpha \, [\mathbb{O}^{\gamma_T^\mu\gamma^5}(\bar \alpha z,0)]_{\text{tw2}}.
\end{align}
Twist-three contributions:
\begin{eqnarray}
\label{OPE:twist3}
\lefteqn{\widetilde C_-\otimes \mathbb{T}^{\mu}_-(z,0) + \widetilde C_+\otimes \mathbb{T}^{\mu}_+(z,0) =}
\nonumber\\
&=& \widetilde C_T^{(1)}\otimes [\mathbb{O}^{\gamma_T^\mu \gamma^5}(z,0)]_{\text{tw3}}
+ \int_0^1 d\alpha \biggl\{
\int_\alpha^1 d\beta \biggl[\frac{N_c}{2} \(\frac{\beta^{2\epsilon}-\beta}{1-2\epsilon}-\alpha\bar \beta (1-\epsilon)\)
\notag\\&&{}
+\frac{1}{2N_c}\frac{\alpha}{2}(2-\alpha+\alpha \epsilon)\biggr]
\Big[\mathbb{T}^\mu_-(z,\beta z,\alpha z) - \mathbb{T}^\mu_+(\bar \alpha z,\bar \beta z,0)\Big]
\nonumber\\
&&{}+\frac{1}{2N_c}\int_0^\alpha d\beta \biggl[
 \frac{\beta}{2}(2-\beta+\beta \epsilon)
\Big[\mathbb{T}^\mu_-(\bar \alpha z,\bar \beta z,0)- \mathbb{T}^\mu_+(z,\beta z,\alpha z)\Big]
\nonumber\\
&&{} +\frac{\beta}{2}\(\frac{2(\alpha^{2\epsilon}-\alpha)}{(1-2\epsilon)\alpha}+\beta+(2-\beta) \epsilon\)
\Big[\mathbb{T}^\mu_-(z,\beta z,\alpha z) - \mathbb{T}^\mu_+(\bar \alpha z,\bar \beta z,0)\Big]
\biggr]
\biggr\},
\end{eqnarray}
where $[\mathbb{O}^{\gamma_T^\mu \gamma^5}(z,0)]_{\text{tw3}}$ is given by Eq.~\eqref{t=3}. Note that the contributions in the first two and the last two lines in Eq.~\eqref{OPE:twist3} correspond to different
ordering of the fields on the light cone: In the first two lines the gluon field 
is in between the quark and the antiquark, and in the last two lines the gluon is outside (to the left or to the right). 
The ``wrongly ordered'' contributions are suppressed in the large-$N_c$ limit.

Renormalized coefficient functions are obtained as explained above, applying the renormalization factors for the light-ray
operators and the overall renormalization factor $Z_{O}$ for the qPDF operator.  
The resulting expression is finite at $\epsilon\to 0$, as it should. 
Cancellation of the $1/\epsilon$ poles provides one with a further check of the 
calculation.

\subsection{qITDs at NLO}
\label{NLOqITDs}

The qITDs are obtained by taking the matrix element of the OPE \eqref{ren:genstr2}
such that the structure of the expansion is essentially retained. 
In the following expressions
\begin{eqnarray}
\mathrm{L}_z=\ln\(\frac{-z^2 \mu^2}{4\,e^{-2\gamma_E}}\),
\end{eqnarray}
and the plus-distribution is defined as usual,
\begin{eqnarray}
\int_0^1 d\alpha\, f(\alpha)\big(g(\alpha)\big)_+=\int_0^1 d\alpha\, \big(f(\alpha)-f(1)\big)g(\alpha)\,.
\end{eqnarray}
We obtain
\begin{align}\label{NLO:G1}
\mathcal{G}_1(\zeta, z^2;\mu) &=\widehat\he(\zeta;\mu)+ a_s\int_0^1 d\alpha \,\mathbf{C}^{(1)}_1(\alpha,\mathrm{L}_z;\mu)\,\widehat\he(\alpha\zeta;\mu)\,,
\end{align}
where
\begin{align}\label{NLO:C1-coef}
\mathbf{C}_1^{(1)}(\alpha,\mathrm{L}_z)= 2 C_F\[\(-\mathrm{L}_z \frac{1+\alpha^2}{1-\alpha}+
\frac{3-8\alpha+3\alpha^2-4\ln \bar\alpha}{1-\alpha}\)_++\delta(\bar \alpha)\(\frac{3}{2}\mathrm{L}_z+\frac{7}{2}\)\].
\end{align}
This result is in agreement with earlier calculations~\cite{Radyushkin:2017lvu,Braun:2018brg, Izubuchi:2018srq}. 
Note, that the coefficient function (\ref{NLO:C1-coef}) is nothing but the renormalized  $\widetilde C_1^{(1)}$ given in Eq.~\eqref{OPE:twist2}.

The transverse qITD $\mathcal{G}_T$ receives twist-two and twist-three contributions
\begin{align}
 \mathcal{G}_T &= \mathcal{G}^{\text{tw2}}_T+ \mathcal{G}^{\text{tw3}}_T,
\end{align}
where the twist-two term is given in Eq.~(\ref{NLO:GT-tw2}) and the twist-three term is given in Eq.~(\ref{NLO:GT}).

First, we present the twist-two part, which reads
\begin{align}\label{NLO:GT-tw2}
\mathcal{G}_T^{\text{tw2}}(\zeta, z^2;\mu)=\widehat{g}_T^{\text{tw2}}(\zeta;\mu)
+ a_s \int_0^1 d\alpha\, \mathbf{C}_T^{(1)}(\alpha,\mathrm{L}_z;\mu)\,\widehat{g}^{\text{tw2}}_T(\alpha\zeta;\mu)\,,
\end{align}
where 
\begin{align}\label{NLO:C2-coef}
\mathbf{C}_T^{(1)}(\alpha,\mathrm{L}_z;\mu)&= 2 C_F\biggl[\(-\mathrm{L}_z \frac{1+\alpha^2}{1-\alpha}+
\frac{1-4\alpha+\alpha^2-4\ln \bar\alpha}{1-\alpha}\)_++\delta(\bar \alpha)\(\frac{3}{2}\mathrm{L}_z+\frac{5}{2}\)\biggr],
\end{align}
and we remind that \eqref{gT-tw2}
\begin{align}
\widehat{g}^{\text{tw2}}_T(\zeta)&=\int_0^1 d\alpha\, \widehat\he(\alpha \zeta). 
\end{align}
Importantly, $\mathbf{C}_T^{(1)}(\alpha,\mathrm{L}_z) \neq \mathbf{C}_1^{(1)}(\alpha,\mathrm{L}_z)$. As a consequence, the 
Wandzura-Wilczek relation does not hold for quasidistributions,
\begin{eqnarray}
\lefteqn{\mathcal{G}^{\text{tw2}}_T(\zeta,z^2) - \int_0^1d\alpha\, \mathcal{G}_1(\alpha \zeta,\alpha^2 z^2) =}
\nonumber\\&=& 8 a_s C_F \int_0^1\!d\alpha\(\Li_2(\bar \alpha)+\ln \bar \alpha \ln \alpha -\frac{\ln^2\alpha}{4}
\)\widehat{\he}\(\alpha \zeta\) + \mathcal{O}(a^2_s)\,.
\label{WWviolation-ITD}
\end{eqnarray}
This is different from the ``classical'' Wandzura-Wilczek relation for polarized DIS \cite{Wandzura:1977qf}, 
which is exact at the level of structure functions, cf.~\cite{Braun:2001qx}.

The twist-three part deserves more attention. 
The first term in Eq.~\eqref{OPE:twist3} is written in terms of the two-particle twist-three operator
$[\mathbb{O}^{\gamma_T^\mu \gamma^5}(z,0)]_{\text{tw3}}$ with the same coefficient function as the twist-two contribution. 
If this ``two-particle'' term were the only one, the qITD $\mathcal{G}_T$ would be factorizable in terms of the 
``full'' PDF\, $ \widehat{g}_T(\zeta)= \widehat{g}^{\text{tw2}}_T(\zeta)+ \widehat{g}^{\text{tw3}}_T(\zeta)$, as 
conjectured in Refs.~\cite{Bhattacharya:2020cen,Bhattacharya:2020xlt}. 
To clarify the role of the remaining three-particle contributions it is instructive 
to consider the large-$N_c$ limit such that the contributions in the last three lines in \eqref{OPE:twist3}) 
can be dropped and the expressions become much simpler. 
We have
\begin{align}
 \mathcal{G}^{\text{tw3}}_T &= \widehat{g}^{\text{tw3}}_T(\zeta;\mu) + 
 a_s \int_0^1 d\alpha\, \mathbf{C}_T^{(1)}(\alpha,\mathrm{L}_z;\mu)\,\widehat{g}^{\text{tw3}}_T(\alpha\zeta;\mu)
\notag\\&\quad
+
2a_s N_c \zeta^2  \int_0^1 d\alpha \int_\alpha^1 d\beta\,
\Big[
   \bar \beta \Big(\bar\alpha\mathrm{L}_z + 2+\alpha\Big) + 2 \ln\beta \Big]
\widehat S^-(\zeta, \beta\zeta,\alpha\zeta) + \mathcal{O}(1/N_c)
\notag\\&=
\widehat{g}^{\text{tw3}}_T(\zeta;\mu) + 
 a_s \int_0^1 d\alpha\, 
\Big[\mathbf{C}_T^{(1)}(\alpha,\mathrm{L}_z;\mu) + N_c\Big(\mathrm{L}_z\(\delta(\bar \alpha)-\alpha\)+\alpha+2\delta(\bar \alpha)\Big)
\Big]\widehat{g}^{\text{tw3}}_T(\alpha\zeta;\mu)
\notag\\&\quad\label{NLO:large-nc}
+
4 a_s N_c \zeta^2  \int_0^1 d\alpha \int_\alpha^1 d\beta\,\ln\beta \,
\widehat S^-(\zeta, \beta\zeta,\alpha\zeta) + \mathcal{O}(1/N_c).
\end{align}
To simplify this expression we have used  Eq.~\eqref{gT-tw3-simpl} and the symmetry relation (\ref{def:S-sym}) which equalizes 
the contribution of $\mathbb{T}_+^\mu$ and $\mathbb{T}_-^\mu$. 

We see that the three-particle contributions involving the  weight factor $\bar\beta$ in the integral over the gluon position 
can be rewritten in terms of the two-particle twist-three PDF $\widehat{g}^{\text{tw3}}_T(\alpha\zeta;\mu)$. However: 
\begin{enumerate}
\item With this addition, the coefficient function  of $\widehat{g}^{\text{tw3}}_T(\alpha\zeta;\mu)$ becomes different 
  from the coefficient function of  $\widehat{g}^{\text{tw2}}_T(\alpha\zeta;\mu)$. 
 This is already true for the logarithmic term $\sim\mathrm{L}_z$, i.e. for 
the evolution kernel in the large-$N_c$ limit~\cite{Ali:1991em,Braun:2001qx}.%
\footnote{Simplification of the twist-three LO evolution equation in the large-$N_c$ limit is due to 
a ``hidden'' symmetry of QCD known as complete integrability~\cite{Belitsky:2004cz}.}
The main effect to the LO accuracy is the constant shift by $N_c$ of the anomalous dimensions 
of the twist-three operators as  compared to the leading twist ones~\cite{Ali:1991em}.        
\item A genuine three-particle contribution involving $\ln \beta$ appears, see the last line of Eq.~(\ref{NLO:large-nc}). It
cannot be rewritten in terms of $\widehat{g}^{\text{tw3}}_T(\alpha\zeta;\mu)$. 
\end{enumerate}
The $1/N_c$ corrections have a more complicated structure and do not allow a separation of two-particle 
contributions in a natural way. Collecting all terms we obtain the NLO expression for the twist-three part 
of the qITD $\mathcal{G}_T(\zeta,z^2)$ (cf.~\eqref{gT-factor}) 
\begin{align}
\label{NLO:GT}
\mathcal{G}_T^{\mathrm{tw3}}(\zeta,z^2)& =  \widehat{g}_T^{\mathrm{tw3}}(\zeta;\mu)
+ 
a_s \mathbf{C}^{(1)}_{\mathrm{2pt}} \otimes \widehat{g}_T^{\mathrm{tw3}}
+ 2\zeta^2a_s \mathbf{C}^{(1)}_{\mathrm{3pt}}\otimes \widehat{S}^- 
\end{align}
with 
\begin{align}
\label{NLO:C2pt}
 \mathbf{C}^{(1)}_{\mathrm{2pt}} \otimes \widehat{g}_T^{\mathrm{tw3}} &=
\int_0^1 d\alpha\,\Big[\mathbf{C}^{(1)}_{T}(\alpha,\mathrm{L}_z;\mu)
+ N_c \Big(\mathrm{L}_z\(\delta(\bar \alpha)-\alpha\)+\alpha+2\delta(\bar \alpha)\Big)
\Big]\,\widehat{g}_T^{\mathrm{tw3}}(\alpha \zeta;\mu)\,,
\\
\label{NLO:C3pt}
\mathbf{C}^{(1)}_{\mathrm{3pt}}\otimes \widehat{S}^- &=
 -\mathrm{L}_z\, \mathrm{P}_{\text{tw3}}\otimes \widehat{S}^-
+\int_0^1 d\alpha\biggl\{\int_\alpha^1 \! d\beta \(2N_c\ln \beta+\frac{1}{N_c}\frac{\alpha^2}{2}\) \widehat{S}^-(\zeta,\beta \zeta,\alpha\zeta)
\notag\\ &\quad
+\frac{1}{N_c}\int_0^\alpha\! d\beta \biggl[\frac{\beta^2}{2}
\widehat{S}^-(\bar \alpha \zeta,\bar \beta \zeta,0)
-\Big(\frac{\beta(2+\beta)}{2}-\frac{2\beta}{\alpha}(1+\ln\alpha)\Big)\widehat{S}^-(\zeta,\beta \zeta,\alpha \zeta)\biggr]\biggr\},
\end{align}
where the logarithmic part is given by
\begin{align}\label{NLO:Ptw3-coef}
\mathrm{P}_{\text{tw3}}\otimes  \widehat{S}^- &=
\frac{1}{N_c}\int_0^1 d\alpha \biggl\{\int_\alpha^1 d\beta \frac{\alpha(\alpha-2)}{2}
\widehat{S}^-(\zeta,\beta \zeta,\alpha\zeta)
\notag\\&\quad
+\int_0^\alpha d\beta \biggl[
\frac{\beta(\beta-2)}{2}\widehat{S}^-(\bar \alpha \zeta,\bar \beta \zeta,0)
+\Big(\frac{\beta(2-\beta)}{2}-\frac{\beta}{\alpha}\Big)\widehat{S}^-(\zeta,\beta \zeta,\alpha \zeta)\biggr]\biggr\}.
\end{align}

\section{Quasi- and pseudo-distributions}
\label{sec:qPDF+pPDF}

qPDFs and pPDFs~\cite{Ji:2013dva,Radyushkin:2016hsy,Izubuchi:2018srq} are defined as functions of the parton momentum fraction
so that their interpretation is more close to the traditional PDFs. There seems to be no established notation in the literature
for all their variants.  Traditionally, the distributions and structure functions related to axial-vector operators 
are denoted by the letter $g$ with various subscripts, cf.~\cite{Kodaira:1978sh,Shuryak:1981pi,Jaffe:1996zw}.
To follow this practice, on the one side, and to distinguish various types of distributions, on the other side, 
we use different fonts: The qPDFs are denoted by the typewriter font letter $\mathtt{g}$, and pPDFs are denoted by the blackletter font 
$\mathfrak{g}$, and similarly the corresponding coefficient functions.

Following \cite{Ji:2013dva,Izubuchi:2018srq} we introduce qPDFs as Fourier transforms of the qITDs 
with respect to the distance $z$. The orientation of the vector $z^\mu$ is kept fixed. We define
\begin{eqnarray}\label{def:quasi1}
\mathtt{g}_1(x,p_v)&=&p_v\int \frac{dz}{2\pi}e^{-ixzp_v}\mathcal{G}_1(zp_v,z^2),
\\\label{def:quasiT}
\mathtt{g}_T(x,p_v)&=&p_v\int \frac{dz}{2\pi}e^{-ixzp_v}\mathcal{G}_T(zp_v,z^2),
\end{eqnarray}
where $v^\mu=z^\mu/|z|$ is the unit vector along $z^\mu$, and $p_v=(p \cdot v)$. 
In turn, pPDFs~\cite{Radyushkin:2017cyf} are defined as Fourier transforms of qITDs with respect to the momentum $p$ keeping its orientation fixed,
\begin{eqnarray}\label{def:pseudo1}
\mathfrak{g}_1(x,z^2)&=&\int \frac{d\zeta}{2\pi}e^{-ix\zeta}\mathcal{G}_1(\zeta,z^2),
\\\label{def:pseudoT}
\mathfrak{g}_T(x,z^2)&=&\int \frac{d\zeta}{2\pi}e^{-ix\zeta}\mathcal{G}_T(\zeta,z^2).
\end{eqnarray}
These distributions have different properties. 
In particular, pseudodistributions have a natural ``partonic'' support $|x|<1$~\cite{Radyushkin:2016hsy}, whereas qPDFs have $|x|<\infty$.

In this section, we present the NLO expressions for axial-vector pPDFs and qPDFs. 
Going over from the qITDs \eqref{NLO:G1}, \eqref{NLO:GT} to the pseudodistributions is relatively straightforward, 
since all Fourier integrals are reduced to Dirac delta-functions and are easily taken. 
The derivation of qPDFs is less trivial due to Fourier transformation of logarithmic contributions. 
These terms have ``unnatural'' support properties, which makes the direct calculation cumbersome. 
To avoid this complication we use the identity
\begin{eqnarray}\label{def:pseudo->quasi}
\mathtt{g}_i(x,p_v)=\int \frac{d\zeta}{2\pi}\int_{-1}^1 dy~e^{i(y-x)\zeta}\mathfrak{g}_i\(y,\frac{\zeta^2}{p_v^2}\), \qquad i=1,T\,,
\end{eqnarray}
from which simple relations between the coefficient functions for qPDF and pPDFs can be derived, see App.~\ref{app:relation}.
Using \eqref{app:p->qn0} \eqref{app:p->qn1} we are able to obtain the NLO expressions for qPDFs from the 
corresponding pPDFs with relatively little effort.

\subsection{pPDFs at NLO}

Using Eqs.~\eqref{NLO:G1}, \eqref{NLO:GT} and performing the Fourier transformation \eqref{def:pseudo1}, \eqref{def:pseudoT} we obtain
\begin{align}\label{pPDF:g1}
\mathfrak{g}_1(x,z^2)& =\he(x)+ a_s \int_{|x|}^1\frac{d\alpha}{\alpha} \mathfrak{C}^{(1)}_1(\alpha,\mathrm{L}_z)\he\Big(\frac{x}{\alpha}\Big)\,,
\\
\mathfrak{g}_T(x,z^2)& =g_T(x)+ a_s
\int_{|x|}^1\frac{d\alpha}{\alpha} \biggl(\mathfrak{C}_T(\alpha,\mathrm{L}_z)g_T^{\text{tw2}}\Big(\frac{x}{\alpha}\Big)
+\mathfrak{C}^{(1)}_{\mathrm{2pt}}(\alpha,\mathrm{L}_z)g_T^{\text{tw3}}\Big(\frac{x}{\alpha}\Big)\biggr)
\notag\\& \hspace*{1.5cm}
+  2 a_s \mathfrak{C}^{(1)}_{\mathrm{3pt}}\otimes {S}^-, 
\label{pPDF:gT}
\end{align}
where
\begin{align}
 \mathfrak{C}^{(1)}_1(\alpha,\mathrm{L}_z) &=  \mathbf{C}^{(1)}_1(\alpha,\mathrm{L}_z)\,, \qquad\text{Eq.}~\eqref{NLO:C1-coef}\,, 
\notag\\
 \mathfrak{C}^{(1)}_T(\alpha,\mathrm{L}_z) &=  \mathbf{C}^{(1)}_T(\alpha,\mathrm{L}_z)\,, \qquad\text{Eq.}~\eqref{NLO:C2-coef}\,, 
\notag\\
 \mathfrak{C}^{(1)}_{\mathrm{2pt}}(\alpha,\mathrm{L}_z) &=  \mathbf{C}^{(1)}_{\mathrm{2pt}}(\alpha,\mathrm{L}_z)\,, \qquad\text{Eq.}~\eqref{NLO:C2pt}\,, 
\end{align}
and
\begin{align}\label{pPDF:R}
\mathfrak{C}^{(1)}_{\mathrm{3pt}}\otimes {S}^- & =-\mathrm{L}_z \mathfrak{P}_{\text{tw3}}\otimes S^-
\notag \\&\quad
+ \int [dx]\int_0^1\! d\alpha\, \Biggl\{\frac{-2N_c}{1-\alpha}\(\frac{\delta(x+\alpha x_1)}{x_1x_3}+\frac{\delta(x-x_3-\alpha x_2)}{x_2x_3}+\frac{\delta(x+x_1)}{x_1x_2}\)
\notag \\&\quad
+\frac{1}{N_c}\biggl[
-\frac{2}{(1-\alpha)_+}\frac{\delta(x+\alpha x_1)}{x_1x_2}
+\frac{2(1+\ln\bar \alpha)}{1-\alpha}\frac{\delta(x+\alpha x_1)-\delta(x+x_1+\bar \alpha x_3)}{x_2^2}
\notag \\&\quad
- \bar \alpha\(
\frac{\delta(x-\alpha x_3)}{x_1x_2}
+\frac{\delta(x-\alpha x_2)}{x_2x_3}
-\frac{\delta(x+\alpha x_2)}{x_1x_2}
-\frac{\delta(x+\alpha x_1)}{x_1x_2}
+\frac{\delta(x+\alpha x_1)}{x_1x_3}\)
\notag \\&\quad
-\(\frac{\delta(x-\alpha x_2)}{x_2x_3}+\frac{\delta(x+\alpha x_1)}{x_1x_3}+2\frac{\delta(x+x_1)}{x_1x_2}\)\biggr]\Biggr\}S^-(x_1,x_2,x_3)\,,
\end{align}
with the logarithmic part being
\begin{align}
\mathfrak{P}_{\text{tw3}}\otimes S^- \!& =\frac{1}{N_c} \int [dx]\int_0^1 d\alpha \Biggl[
\alpha\biggl(
\frac{\delta(x-\alpha x_3)}{x_1x_2}+\frac{\delta(x-\alpha x_2)}{x_2x_3}
-\frac{\delta(x+\alpha x_2)}{x_1x_2}
-\frac{\delta(x+\alpha x_1)}{x_1x_2} 
\notag\\&
+\frac{\delta(x+\alpha x_1)}{x_1x_3}+2\frac{\delta(x+x_1)}{x_1x_2}\biggr)
+ \frac{\delta(x\!+\!x_1\!+\!\bar \alpha x_3)-\delta(x\!+\!\alpha x_1)}{(1-\alpha)x_2^2}\Biggr]S^-(x_1,x_2,x_3).
\label{pPDF:P}
\end{align}
One can show that the integrands in these expressions are finite at $x_i\to0$ and $\alpha\to1$. 
Moreover, the first and the second moments of (\ref{pPDF:R}) and (\ref{pPDF:P}) vanish. All expressions are defined for $-1<x<1$. 

\subsection{qPDFs at NLO}

The qPDFs can be most easily derived  from the corresponding pPDFs \eqref{pPDF:g1}, \eqref{pPDF:gT} applying the double-Fourier transformation \eqref{def:pseudo->quasi}.
This transformation is non-trivial only for the terms involving $\mathrm{L}_z$ \eqref{app:p->qn1},
which are also the ones responsible for extending the support property of qPDFs 
beyond the partonic region $|x|<1$ to $|x|<\infty$. In the expressions given below we use the following notation: 
\begin{align}
\(f(x)\)_\oplus&= f(x)-\delta(\bar x)\int_1^\infty\! f(x)\,dx,
\\
\(f(x)\)_\ominus&= f(x)-\delta(x)\int_{-\infty}^0\! f(x)\,dx.
\end{align}
The first moment of both distributions is zero. 

We obtain
\begin{align}
\mathtt{g}_1(x,p_v)& =\he(x)+ a_s \int_{-1}^1\frac{dy}{|y|} \mathtt{C}^{(1)}_1\Big(\frac{x}{y},\mathrm{L}_p\Big)\,\he(y)\,,
\label{qPDF:g1}\\
\mathtt{g}_T(x,p_v)& =g_T(x)+ a_s 
\int_{-1}^1\frac{dy}{|y|} \biggl(\mathtt{C}^{(1)}_T\Big(\frac{x}{y},\mathrm{L}_p\Big)g_T^{\text{tw2}}(y)
+\mathtt{C}^{(1)}_{\mathrm{2pt}}\Big(\frac{x}{y},\mathrm{L}_p\Big) g_T^{\text{tw3}}(y)\biggr)
\notag\\& \hspace*{1.5cm}
+  2 a_s \mathtt{C}^{(1)}_{\mathrm{3pt}}\otimes {S}^-, 
\label{qPDF:gT}
\end{align}
where
\begin{align}
\mathrm{L}_p& =\ln\(\frac{\mu^2}{4 y^2 p_v^2}\),
\end{align}
and the coefficient functions are given by
\begin{eqnarray}\label{qPDF:C1}
\texttt{C}_1^{(1)}(\alpha,\mathrm{L}_p)&=& 2 C_F \delta(\bar \alpha)\(\frac{3}{2}\mathrm{L}_p+\frac{7}{2}\)
\\\nn &&+2 C_F\left\{
\begin{array}{lc}
\Ds \(\frac{1+\alpha^2}{1-\alpha}\ln\frac{\alpha}{-\bar \alpha}+1-\frac{3}{2\bar \alpha}\)_\oplus,&\alpha>1
\\
\Ds \(\frac{1+\alpha^2}{1-\alpha}\[-\mathrm{L}_p +\ln(\alpha\bar \alpha)-1\]+3-2\alpha+\frac{3}{2\bar \alpha}\)_+,&0<\alpha<1
\\
\Ds \(\frac{1+\alpha^2}{1-\alpha}\ln\frac{\bar \alpha}{-\alpha}-1+\frac{3}{2\bar \alpha}\)_\ominus,&\alpha<0
\end{array}
\right.
\\
\label{qPDF:CT}
\mathtt{C}_T^{(1)}(\alpha,\mathrm{L}_p)&=&2 C_F\delta(\bar \alpha)\(\frac{3}{2}\mathrm{L}_p+\frac{5}{2}\)
\\\nn &&+2 C_F\left\{
\begin{array}{lc}
\Ds \(\frac{1+\alpha^2}{1-\alpha}\ln\frac{\alpha}{-\bar \alpha}+1-\frac{3}{2\bar \alpha}\)_\oplus,&\alpha>1
\\
\Ds \(\frac{1+\alpha^2}{1-\alpha}\[-\mathrm{L}_p +\ln(\alpha\bar \alpha)-1\]+1+\frac{3}{2\bar \alpha}\)_+,&0<\alpha<1
\\
\Ds \(\frac{1+\alpha^2}{1-\alpha}\ln\frac{\bar\alpha}{-\alpha}-1+\frac{3}{2\bar \alpha}\)_\ominus,&\alpha<0
\end{array}
\right.
\\
\mathtt{C}_{\mathrm{2pt}}^{(1)}(\alpha,\mathrm{L}_p)&=&\mathtt{C}_T^{(1)}(\alpha,\mathrm{L}_p)
+ N_c\[\delta(\bar \alpha)\({\frac{1}{2}}\mathrm{L}_p+\frac{5}{2}\)
+\theta(0\!<\!\alpha\!<\!1){[\alpha(1-\rm{L_p})]_+}+\mathtt{r}_1(\alpha)\]\, ,\phantom{aaaaa}
\label{qPDF:C2pt}
\end{eqnarray}
and
\begin{align}
\mathtt{C}^{(1)}_{\mathrm{3pt}}\otimes {S}^- &= \theta(|x|<1)\biggl[\mathfrak{C}^{(1)}_{\mathrm{3pt}}\otimes {S}^-{\Big|_{{\rm L}_z\mapsto{\rm L}_p}}
\notag\\&\quad
- \frac{1}{2N_c} \mathrm{L}_p \int [dx] \biggl(\frac{\delta(x\!-\!x_3)-\delta(x\!+\!x_2)}{x_1x_2}
-\frac{\delta(x\!+\!x_1)-\delta(x\!-\!x_2)}{x_2x_3}\biggr)S^-(x_1,x_2,x_3)\biggr]
\notag\\ &\quad
+\frac{1}{N_c} \int [dx]\int_{-\infty}^\infty d\alpha \biggl[
\mathtt{r}_1(\alpha)\biggl(\frac{\delta(x-\alpha x_3)}{x_1x_2}
+\frac{\delta(x-\alpha x_2)}{x_2x_3}
-\frac{\delta(x+\alpha x_2)}{x_1x_2}
\notag\\ &\qquad
-\frac{\delta(x+\alpha x_1)}{x_1x_2}
 +\frac{\delta(x+\alpha x_1)}{x_1x_3}
+2\frac{\delta(x+x_1)}{x_1x_2}\biggr)
\notag\\ &\qquad
+\mathtt{r}_2(\alpha)\frac{\delta(x+x_1+\bar \alpha x_3)-\delta(x+\alpha x_1)}{x_2^2}\biggr]S^-(x_1,x_2,x_3)\,,
\label{qPDF:R}
\end{align}
where
\begin{align}
\mathtt{r}_1(\alpha)&=
\left\{
\begin{array}{lc}
\Ds \(\alpha \ln\frac{\alpha}{-\bar \alpha}-1+\frac{1}{2\bar \alpha}\)_\oplus,&\alpha>1
\\
\Ds \(1-2\alpha+\alpha \ln(\alpha\bar \alpha)-\frac{1}{2\bar \alpha}\)_+,&0<\alpha<1
\\
\Ds \(\alpha \ln\frac{\bar \alpha}{-\alpha}+1-\frac{1}{2\bar \alpha}\)_\ominus,&\alpha<0
\end{array}\right.
\\[2mm]
\mathtt{r}_2(\alpha)&=
\left\{
\begin{array}{lc}
\Ds \(\frac{\ln\frac{\alpha}{-\bar \alpha}}{1-\alpha}\)_\oplus,&\alpha>1
\\
\Ds \(\frac{3\ln \bar \alpha+\ln \alpha}{1-\alpha}\)_+,&0<\alpha<1
\\
\Ds \(\frac{\ln\frac{\bar \alpha}{-\alpha}}{1-\alpha}\)_\ominus,&\alpha<0\,.
\end{array}\right.
\end{align}
Note that in the large-$N_c$ limit the genuine three-particle contributions for qPDFs and pPDFs are the same.

Our results for the coefficient functions $\texttt{C}_1$ and $\texttt{C}_T$ coincide with known expressions~\cite{Radyushkin:2017lvu,Izubuchi:2018srq}.
The remaining expressions are new results. 
In general, the structure of the three-particle contributions to qPDFs is rather unwieldy due to multitude of integration regions.
For two-particle contributions, as well known, three domains  $x>1$, $0<x<1$ and $x<0$ are distinguished.
For three-particle contributions one ends up with 30 domains for the variables $(x_1,x_2,x_3)$. This structure is not explicit in Eq.~(\ref{qPDF:R})
but is revealed once the integrations over the delta-functions are performed.

\section{Discussion and outlook}

We have formulated the factorization theorem for the space-like axial-vector correlation function (\ref{def:op-main}) in terms
of parton distributions to twist-three accuracy and calculated the corresponding coefficient functions to NLO accuracy. In this
section we discuss the results in connection with possible lattice calculations.

Since the twist-two contributions to the ``transverse'' part of all versions of the quasi-parton distributions are given by 
the helicity PDF $\he(x,\mu)$, the utility of the lattice approach crucially depends on the 
possibility to subtract (or at least minimize) such terms and reveal the twist-three contributions of interest. 
It is well-known that this subtraction can be implemented exactly for the correlation function of two vector currents by virtue of the Wandzura-Wilczek 
relation. For the quasidistributions the cancellation is not complete, 
as explicitly demonstrated by our calculation. We obtain
\begin{align}\label{residual1}
&\mathtt{g}_T(x,p_v) - \int_{|x|}^1\frac{dy}{y} \mathtt{g}_1(y,p_v) =
\notag\\ & \hspace*{1cm} =
 8a_sC_F \int_{|x|}^1 \frac{dy}{y}
\(\Li_2(\bar y)+\ln \bar y \ln y -\frac{\ln^2y}{4}
\)\he\(\frac{x}{y}\) + \text{twist~three},
\\\label{residual2}
& \mathfrak{g}_T(x,z^2) - \int_{|x|}^1\frac{dy}{y} \mathfrak{g}_1(y,z^2)
~=~
 4a_sC_F \int_{|x|}^1 \frac{dy}{y}\(\bar y+\ln y\)\he\(\frac{x}{y}\) + \text{twist~three}\,,
\end{align} 
for the qPDFs and pPDFs, respectively.
The twist-two remainder on the r.h.s. of this relation for the qPDF case is unfortunately rather large.

To illustrate this point, consider the first nontrivial moment of $g_T$ which can be accessed by 
considering the small-distance expansion of the qITD. 
Using Eqs.~\eqref{WWviolation-ITD} and \eqref{NLO:GT} one obtains
\begin{align}
&\mathcal{G}_T(\zeta,z^2) - \int_0^1d\alpha\, \mathcal{G}_1(\alpha \zeta,\alpha^2 z^2) 
=
 4a_sC_F\,a_0+i\zeta \frac52 a_sC_F\, a_1
\\\nn &
+\frac{\zeta^2}{3}\Bigg\{ \widetilde{d}_2
-\frac{20}{9}C_Fa_s a_2+a_s\widetilde{d}_2\[L_z\(\frac{13}{3}N_c-\frac{4}{3 N_c}\) - 8 C_F\]\Bigg\}
+\mathcal{O}(\zeta^3,z^2)\, ,
\end{align} 
where  (we use the notations of Ref.~\cite{Armstrong:2018xgk}) 
\begin{align}
 a_n =  \int_{-1}^1dx\, x^n \he(x)\,, && \widetilde d_2 = \int_0^1dx\, x^2[3 g_T(x) - \he(x)] = \int [dx]\, S^-(x_1,x_2,x_3)\,.
\end{align}
For an estimate, we take $a_2 \simeq 0.05$ for the $u$-quarks in the proton at $\mu^2=4~\text{GeV}^2$~\cite{deFlorian:2009vb,Blumlein:2010rn}.
Then $(20/9)C_F a_s a_2 \sim 3.6\cdot 10^{-3}$ which is of the same order as the expected size of the twist-three matrix element 
$|\tilde d_2| \sim (\text{1\textdiv 5})\cdot 10^{-3}$
\cite{Balla:1997hf,Braun:2011aw,Gockeler:2005vw,Armstrong:2018xgk}.    
Reducing this ``twist-two pollution'' can pose a serious problem for the qPDF approach in the studies of 
twist-three effects
\footnote{The residual twist-two contribution for the combination 
$\mathcal{G}_T(\zeta,z^2) - \int_0^1d\alpha\, \mathcal{G}_1(\alpha \zeta, z^2)$ is smaller, compare Eqs.~(\ref{residual1}) and (\ref{residual2}). It is more difficult, however, 
to implement this subtraction in the lattice data.}.

As far as the twist-three contribution itself is concerned, constraining 
the quark-antiquark-gluon correlation function in its full complexity from present-day lattice calculations 
is probably unrealistic. Thus trying to reduce the nonperturbative input 
to a function of one variable, $g_T(x)$, as attempted in~\cite{Bhattacharya:2020cen,Bhattacharya:2020xlt}, 
is certainly logical.
However, the shortcomings of such a reduction have to be clearly understood. Any approximation of this kind 
is theoretically self-consistent if and only if it is maintained at all scales, in other words if $g_T(x)$ does not mix 
with the ``genuine'' three-particle contributions that are neglected. This condition is, indeed, satisfied to LO accuracy in the large $N_c$
limit \cite{Ali:1991em,Braun:2001qx}, which can be sufficient at the 
current stage. 
This decoupling does not mean, however, that the coefficient functions of $g_T(x)$ to logarithmic accuracy 
can be calculated from the two-particle quark-antiquark matrix elements. The quark-antiquark-gluon matrix elements must be 
considered and contribute to the splitting functions (and to finite terms). 
As the result, the coefficient functions of the twist-two and twist-three contributions to  $g_T(x)$ in the factorization theorem for the qPDFs are different already in the 
large $N_c$ limit, see \eqref{qPDF:C2pt}. This difference is missed in ~\cite{Bhattacharya:2020cen,Bhattacharya:2020xlt}.
Another issue is that at NLO finite corrections $\sim N_c$ appear that cannot be 
reduced to $g_T^{\mathrm{tw3}}$. Whether such terms can be minimized in some way, remains to be studied. 

To conclude, we have presented the first NLO analysis of axial-vector quasidistributions of the nucleon 
to the twist-three accuracy.
The same method can be extended in a straightforward manner to chiral-odd twist-three quasidistributions 
that are of particular interest, cf.~\cite{Bhattacharya:2020jfj}. 
We plan to consider them in a separate publication.

\acknowledgments

This study was supported by Deutsche Forschungsgemeinschaft (DFG) through the Research 
Unit FOR 2926, ``Next Generation pQCD for Hadron Structure: Preparing for the EIC'', project number 40824754.
Y.J. also acknowledges the support of DFG grant SFB TRR 257.

\appendix

\section*{Appendices}

\section{Evolution kernel for twist-3 distributions}
\label{app:evolution_kernel}

The evolution equations for twist-three quark-antiquark-gluon distributions can be found in Ref.~\cite{Braun:2009mi,Ji:2014eta}. 
For the readers' convenience, we collect the relevant expressions in this appendix.

The evolution equation for the function $\widehat{S}^-$ has the form 
\begin{eqnarray}
\mu^2\frac{d}{d\mu^2}\widehat{S}^-(z_1,z_2,z_3)=-a_s[\mathbb{H}\otimes \widehat{S}^-](z_1,z_2,z_3),
\end{eqnarray}
where
\begin{align}\label{app:kernel1}
[\mathbb{H}\otimes \widehat{S}](z_1,z_2,z_3)&=
N_c\int_0^1 d\alpha\Big(
\frac{4}{\alpha}\widehat{S}(z_1,z_2,z_3)
-\frac{\bar \alpha}{\alpha} \widehat{S}(z_{12}^\alpha,z_2,z_3)
-\frac{\bar \alpha}{\alpha} \widehat{S}(z_1,z_2,z_{32}^\alpha)
\notag\\&\quad
-\frac{\bar \alpha^2}{\alpha} \widehat{S}(z_1,z_{21}^\alpha,z_3)
-\frac{\bar \alpha^2}{\alpha} \widehat{S}(z_1,z_{23}^\alpha,z_3)
-2\int_0^{\bar \alpha}d\beta\,\bar \beta \widehat{S}(z_1,z_{23}^\beta ,z_{32}^\alpha)\Big)
\notag\\&\quad
-\frac{1}{N_c}\int_0^1 d\alpha \Big(
\frac{2}{\alpha}\widehat{S}(z_1,z_2,z_3)
-\frac{\bar \alpha}{\alpha} \widehat{S}(z_{13}^\alpha,z_2,z_3)
-\frac{\bar \alpha}{\alpha} \widehat{S}(z_1,z_2,z_{31}^\alpha)
\notag\\&\quad
+\bar \alpha \,\widehat{S}(z_2,z_{12}^\alpha,z_3)
-\int_0^{\bar \alpha}d\beta \,\widehat{S}(z_{13}^\alpha,z_2,z_{31}^\beta)
+2\int_{\bar \alpha}^1 d\beta\,\bar \beta \,\widehat{S}(z_1,z_{23}^\beta,z_{32}^\alpha)\Big)
\notag\\&\quad
-3C_F\widehat{S}(z_1,z_2,z_3).
\end{align}
Here 
\begin{eqnarray}
\bar \alpha =1-\alpha,\qquad z_{ij}^\alpha=z_i\bar \alpha+z_j \alpha.
\end{eqnarray}
The evolution equation for $\widehat{S}^+$ is obtained trivially, using the symmetry relation (\ref{def:S-sym}). 
The corresponding expression for momentum space distributions is lengthy as the evolution kernels 
in different sectors $x_i\lessgtr 0$ are not the same. Explicit expression can be found in \cite{Ji:2014eta}.

The distribution $\widehat{g}_T$ is proportional to the integral (\ref{gT-tw3-simpl}).
Application of the operator $\mathbb{H}$ to this expression yields
\begin{eqnarray}
\int_0^1\! d\alpha\! \int_\alpha^1\! d\beta \,\bar \beta\, [\mathbb{H}\otimes \widehat{S}](\zeta,\beta \zeta,\alpha \zeta)
&=&
\int_0^1 d\alpha \int_\alpha^1 d\beta \Big[
N_c \bar \beta (\ln \bar \alpha -2\ln\alpha +1)
\\&&{}\hspace*{-3.5cm}
-\frac{1}{N_c} \(\alpha\beta-\frac{\alpha^2}{2}+\bar \beta(2\ln\alpha-\ln\bar \alpha)\)-3C_F \bar \beta \Big]\widehat{S}(\zeta,\beta \zeta,\alpha \zeta)
\notag\\&&{}\hspace*{-3.5cm}
-\frac{1}{N_c}\int_0^1 d\alpha\int_0^\alpha d\beta \Big[\(\frac{\beta}{\alpha}-\frac{\beta(2-\beta)}{2}\)\widehat{S}(\zeta,\beta \zeta,\alpha \zeta)
+\beta \frac{2-\beta}{2}\widehat{S}(\bar \alpha\zeta,\bar \beta \zeta,0)\Big].
\nonumber
\end{eqnarray}
The $\sim N_c$ part of this expression that includes $\bar \beta f(\alpha)$ in the integrand, can be rewritten as a convolution of a two-particle kernel with $\widehat{g}_T$ with the proper rescaling of $\zeta$. 
The $1/N_c$ contributions cannot be simplified in this way and contribute to the ``genuine'' three-point part, see Sect.~\ref{NLOqITDs}.

\section{Relation between the coefficient functions for qPDFs and pPDFs}
\label{app:relation}

The qPDF $\mathtt{g}(x,p_v)$ is related to the pPDF $\mathfrak{g}(x,z^2)$ 
by the double Fourier transformation
\begin{eqnarray}\label{app:p->q}
\mathtt{g}(x,p_v)=\int \frac{d\zeta}{2\pi}\int_{-1}^1 dy e^{i(y-x)\zeta}\mathfrak{g}\(y,\frac{\zeta^2}{p_v^2}\).
\end{eqnarray}
The pPDF on the r.h.s. of this equation can usually be presented in the form of the Mellin convolution
\begin{eqnarray}\label{app:pseudo}
\mathfrak{g}(x,z^2)=\int_{-1}^1 dy \int_0^1 d\alpha\, \delta(x-\alpha y) \mathfrak{C}\(\alpha,\mathrm{L}_z\)Q(y)+O(z^2),
\end{eqnarray}
where $Q(x)$ (with $-1<x<1$) is some parton distribution function. 
Applying the transformation (\ref{app:p->q}) to (\ref{app:pseudo}) one obtains
\begin{eqnarray}\label{app:quasi}
\mathtt{g}(x,p_v)=\int_{-1}^1 dy \int_{-\infty}^\infty d\alpha \delta(x-\alpha y) \mathtt{C}\(\alpha,\mathrm{L}_p\)Q(y)+O(z^2),
\end{eqnarray}
where the coefficient function $\mathtt{C}(x,\mathrm{L}_p)$ is related to $\mathfrak{C}(\alpha,\mathrm{L}_z)$
as 
\begin{eqnarray}\label{app:p->q-for-coeff}
\mathtt{C}\(x,\mathrm{l}_p\)=\int_0^1 d\alpha \int \frac{d\zeta}{2\pi}e^{i(\alpha-x)\zeta}\mathfrak{C}\(\alpha,\mathrm{L}_z\).
\end{eqnarray}
In these formulas we denote
\begin{eqnarray}
\mathrm{L}_z=\ln\(\frac{-z^2\mu^2}{4e^{-2\gamma_E}}\),\qquad
\mathrm{l}_p=\ln\(\frac{\mu^2}{4p_v^2}\),\qquad
\mathrm{L}_p=\ln\(\frac{\mu^2}{4|y|^2p_v^2}\).
\end{eqnarray}
Let us emphasize the factor $|y|$ in the $\mathrm{L}_p$ that is present in the quasidistribution (\ref{app:quasi}). It appears due to the change of variables. Also we observe that the quasi-distribution is nonvanishing for $-\infty<x<\infty$.

The coefficient function $\mathfrak{C}(\alpha,\mathrm{L}_z)$ depends on $\zeta$ only via the logarithms
which appear in increasing powers in higher orders of the perturbative expansion. 
The nontrivial part of the transformation \eqref{app:p->q-for-coeff} 
consists of the evaluation of the Fourier transform of $\ln^n\zeta$. 
Using that $\mathrm{L}_z=\mathrm{l}_p+\ln(\zeta^2 e^{2\gamma_E})$ we write the coefficient function $\mathfrak{C}$ as
a series
\begin{eqnarray}
\mathfrak{C}\(\alpha,\mathrm{L}_z\)=\sum_{n=0}^\infty \ln^n(\zeta^2e^{2\gamma_E}) \mathfrak{C}_{n}(\alpha,\mathrm{l}_p).
\end{eqnarray}
Here $\mathfrak{C}_n$ is given by a perturbative series starting from $a_s^n$. The coefficient function for the qPDF is
then
\begin{eqnarray}
\mathtt{C}\(\alpha,\mathrm{l}_p\)=\sum_{n=0}^\infty \mathtt{C}_{n}(\alpha,\mathrm{l}_p),
\end{eqnarray}
where 
\begin{eqnarray}\label{app:p->q-for-coeff22}
\mathtt{C}_n\(x,\mathrm{l}_p\)=  \int_0^1 d\alpha\,\mathfrak{C}_{n}(\alpha,\mathrm{l}_p) \int \frac{d\zeta}{2\pi}e^{i(\alpha-x)\zeta}\ln^n(\zeta^2 e^{2\gamma_E}).
\end{eqnarray}

The integral \eqref{app:p->q-for-coeff22} can be evaluated  explicitly for any given $n$. 
To this end one should first move the integration contour over $\alpha$ to the complex plane, 
evaluate the integral over $\zeta$, and finally close the $\alpha$-integration contour 
on the branch cuts of the integrand. 
The result for arbitrary $n$ is simple but lengthy. To the NLO accuracy we only need $n=0$ and $n=1$:
\begin{eqnarray}\label{app:p->qn0}
\mathtt{C}_{0}\(x,\mathrm{L}_p\)=
\left\{\begin{array}{lc}
\Ds 0,& x>1 \text{~or~} x<0,
\\\Ds
\mathfrak{C}_{0}(x,\mathrm{L}_p),& 0<x<1.
\end{array}\right.
\end{eqnarray}
and
\begin{eqnarray}\label{app:p->qn1}
\mathtt{C}_{1}\(x,\mathrm{L}_p\)=
\left\{\begin{array}{lc}
\Ds \int_0^1 dy \frac{\mathfrak{C}_{1}(y,\mathrm{L}_p)}{y-x},&  x>1,
\\
\Ds \int_0^x dy \frac{\mathfrak{C}_{1}(y,\mathrm{L}_p)-\mathfrak{C}_{1}(x,\mathrm{L}_p)}{y-x}
+  \int_x^1 dy \frac{\mathfrak{C}_{1}(y,\mathrm{L}_p)-\mathfrak{C}_{1}(x,\mathrm{L}_p)}{x-y}
\\ \Ds \qquad
 -\ln(x\bar x)\mathfrak{C}_{1}(x,\mathrm{L}_p),&  0<x<1,
\\
\Ds \int_0^1 dy \frac{\mathfrak{C}_{1}(y,\mathrm{L}_p)}{x-y},&  x<0.
\end{array}\right.
\end{eqnarray}

Derivation of Eq.~\eqref{app:p->qn1} is straightforward under the assumption that the integrand is 
regular for $y\in[0,1]$. However, $\mathfrak{C}(y)$  has a singularity at $y\to1$ 
in the form of plus distributions. These singularities result in additional 
pole terms $\sim\delta(\bar x)$ that are easy to miss in a direct evaluation.  
To bypass this complication, we have used that the first Mellin moment of plus distributions 
vanishes and thus these additional contributions can be found by enforcing vanishing of the first 
moment of $\mathtt{C}(x)$ by adding suitable $\sim \delta(\bar x)$ terms.

\section{Sample calculations}
\label{app:diagrams}

For pedagogical purposes we present in detail a part of the calculation.
We used two independent methods which we call \textit{Method A} and \textit{Method B}. 
\textit{Method A} is based on explicit expansion of the background field Lagrangian with subsequent evaluation of the diagrams. 
\textit{Method B} uses partially integrated action with the propagators in the background field. 
Although both methods are based on the same concept,
the actual calculation is rather different and in particular the diagrammatic decomposition of the relevant contributions is not the same, 
although there are correspondences between classes of diagrams. For illustration, in \textit{Method A} and \textit{Method B} 
we also employ a different gauge fixing condition for the background field,  
leading to essentially different algebra and intermediate expressions. Naturally, the final results coincide.

\begin{figure}[t]
\begin{center}
\includegraphics[width= 0.95\textwidth]{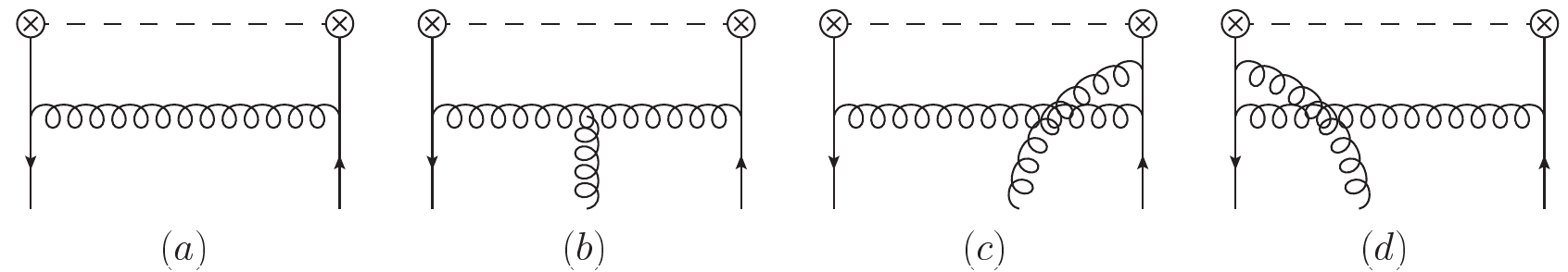}
\end{center}
\caption{\label{fig:diagrams} Diagrams computed in this appendix. The external lines represent the background fields. The crossed blobs connected by the dashed line represent the quasidistribution operator.}
\end{figure}

In this appendix, we present the calculation of the subset of diagrams which involve gluon exchange between the quark and the 
antiquark, see Fig.~\ref{fig:diagrams}. This proves to be the most cumbersome part. In what follows we calculate 
the relevant contributions using both methods and find the same result. 

\subsection{Method A}

Interaction with the background field is described by the action
\begin{eqnarray}
iS_{int}&=&ig\int d^dx\Big[\bar q \slashed{B}\psi+\bar \psi \slashed{B}q+\bar \psi \slashed{A}\psi
+f_{ABC}v^{\mu\alpha\beta \gamma} A_\mu^A(\partial_\alpha B_\beta^B)B_\gamma^C+\ldots\Big],
\end{eqnarray}
where $q$ and $A$ are classical background fields, $\psi$ and $B$ are quantum fields,
$f_{ABC}$ are the $SU(N_c)$ structure constants,  and 
\begin{eqnarray}
v^{\mu\alpha\beta \gamma}=2g^{\mu\beta}g^{\alpha\gamma}-g^{\mu\alpha}g^{\beta\gamma}-2g^{\mu\gamma}g^{\alpha\beta}.
\end{eqnarray}
The dots indicate terms that are not needed in the present computation. 

Using this effective Lagrangian we derive the expressions for the diagrams shown in Fig.~\ref{fig:diagrams}
\footnote{The same calculation can be done in momentum space, leading to identical results.},
\begin{eqnarray}\nn
D_{(a)}&=& (ig)^2 C_F \int d^dx d^dy \, \bar q(x)\gamma^\nu \slashed{\Delta}(x)\gamma^\mu_T\gamma^5 \slashed{\Delta}(z-y)\gamma_\nu q(y)\Delta_g(x-y),
\\\label{app:A:0}
D_{(b)}&=& (ig)^3 \frac{iN_c}{2} \int d^dx d^dy d^du \,\bar q(u) \gamma_\beta \slashed{\Delta}(u)\gamma^\mu_T\gamma^5 \slashed{\Delta}(z-x)\gamma_\gamma q(x)
\\\nn &&\times\(
v^{\nu\alpha\beta \gamma}\partial_\alpha\Delta_g(y-u) \Delta_g(x-y)+v^{\nu\alpha\gamma\beta }\Delta_g(y-u)\partial_\alpha \Delta_g(x-y)\)A_\nu(y),
\\\nn
D_{(c)}&=& -\frac{(ig)^3}{2N_c}\int d^dx d^dy d^du \, \bar q(y)\gamma_\rho \slashed{\Delta}(y-x)\gamma^\nu \slashed{\Delta}(x)\gamma^\mu \gamma^5 \slashed{\Delta}(z-u)\gamma^\rho q(u) A_\nu(x)\Delta_g(u-y),
\\\nn
D_{(d)}&=& -\frac{(ig)^3}{2N_c}\int d^dx d^dy d^du \, \bar q(y)\gamma_\rho \slashed{\Delta}(u)\gamma^\mu \gamma^5 \slashed{\Delta}(z-x)\gamma^\nu \slashed{\Delta}(x-y)\gamma^\rho q(y) A_\nu(x)\Delta_g(u-y),
\end{eqnarray}
where we have already done the color algebra, and
\begin{eqnarray}
\slashed{\Delta}(x)=\frac{\Gamma(2-\epsilon)}{2\pi^{d/2}}\frac{i \slashed x}{(-x^2+i0)^{2-\epsilon}},\qquad
\Delta_g(x)=\frac{\Gamma(1-\epsilon)}{4\pi^{d/2}}\frac{-1}{(-x^2+i0)^{1-\epsilon}}
\end{eqnarray}
are the quark and gluon propagators in position space (with $d=4-2\epsilon$)
in Feynman gauge. 

We use the axial gauge for the background gluon field
\begin{eqnarray}
z^\mu A_\mu(x)=0.
\end{eqnarray}
With this choice that there is no Wilson line, $[z,0] = \II$, and
\begin{align}
  A_\mu(x) = - \int_{-\infty z_0}^0 d\sigma\,  F_{\mu z}(x + \sigma z)\,.
\end{align}
Note that here we use the retarded prescription to fix the residual gauge dependence, $A_\mu(t\to-\infty)=0$. 
In the case of the advanced prescription, $A_\mu(t\to+\infty)=0$, the lower limit of the integration would become $+\infty z_0$.
In the following we tacitly assume $z_0>0$.

The diagrams are to be evaluated in the limit $z^2\to0 $ neglecting terms $\mathcal{O}(z^2)$. 
A similar computation is made in Ref.~\cite{Scimemi:2019gge} for a TMD operator 
which is different from the present case only by $z^\mu\to z^\mu_T$. 
A very detailed description of the calculation of the (analog of) diagram $D_{(d)}$ can be found in App.~B of Ref.~\cite{Scimemi:2019gge}. 
Here we consider $D_{(a)}$, which is algebraically simpler but incorporates all the same principal steps. 
In \textit{Method B}, the same contribution is given by a certain mixture of $D_{(a)}$ and $D_{(b)}$.

Thus we start with (cf. Eq.~(\ref{app:b:1}))
\begin{align}
D_{(a)}&=g^2C_F\frac{\Gamma^2(2-\epsilon)\Gamma(1-\epsilon)}{16\pi^{3d/2}}
\int d^dx d^dy \,\frac{\bar q(x)\gamma^\nu \slashed{x}\gamma^\mu_T\gamma^5 (\slashed{z}-\slashed{y})\gamma_\nu q(y)}{[-x^2]^{2-\epsilon}[-(z-y)^2]^{2-\epsilon}[-(x-y)^2]^{1-\epsilon}}.
\end{align}
The expansion of the integral in powers of $z^2$ can be done with the following trick. 
We join the denominators using auxiliary integrations with Feynman parameters, and perform a shift of integration variables such that the denominator acquires 
the form $A+z^2 B$ where $A$ does not depend on $z$. We obtain
\begin{align}\label{app:A:1}
D_{(a)}&=g^2C_F\frac{\Gamma(5-3\epsilon)}{16\pi^{3d/2}}\int_0^1 [d\alpha d\beta d\gamma] \alpha^{1-\epsilon}\beta^{1-\epsilon}\gamma^{-\epsilon}
\int d^dx d^dy \,
\\\nn &{}\times 
\bar q\(x+\frac{\beta \gamma}{\lambda}z\)
\frac{\gamma^\nu \(\slashed{x}+\frac{\beta\gamma}{\lambda}\slashed{z}\)
\gamma^\mu_T\gamma^5 
\(\frac{\alpha\gamma}{\lambda}\slashed{z}-\slashed{y}\)\gamma_\nu }{[-\bar \beta x^2-\bar \alpha y^2+2\gamma(xy)-\frac{\alpha\beta\gamma}{\lambda}z^2]^{5-3\epsilon}}
q\(y+\frac{\alpha \beta+\beta \gamma}{\lambda}z\),
\end{align}
where $\lambda=\alpha\beta+\alpha\gamma+\beta\gamma$ and 
$[d\alpha d\beta d\gamma]=d\alpha d\beta d\gamma\delta(\alpha+\beta+\gamma-1)$. Next, we expand the quark fields at $x,y\to 0$. 
The resulting integrals are of the form $x^{\mu_1}...y^{\mu_n}/[A+z^2 B]^{5-3\epsilon}$ and can be easily taken. 

To the required twist-three accuracy the diagram should be evaluated up to terms $\sim\bar q\partial q$ so that we need to 
expand the fields in Eq.~(\ref{app:A:1}) to first order. Computing the (tadpole) loop integrals one ends up with 
\begin{align}\label{app:A:2}
D_{(a)}&=(1-\epsilon)\mathbb{N}C_F\int_0^1 [d\alpha d\beta d\rho]
q(\bar \alpha z)\Big[2\gamma^\mu_T-{\rho}(\gamma^\mu_T \slashed{z}\overrightarrow{\slashed{\partial}}-\overleftarrow{\slashed{\partial}}\slashed{z}\gamma^\mu_T)\Big]\gamma^5q(\beta z)+ \mathcal{O}(z^2)\,,
\end{align}
where 
\begin{eqnarray}\label{app:A:N}
\mathbb{N}=\frac{g^2}{16\pi^{d/2}}\Gamma(-\epsilon)(-z^2 \mu^2)^{\epsilon}.
\end{eqnarray}
To present the expression (\ref{app:A:2}) in this simple form we have used 
that $z^\mu_T=0$ and $\{\gamma^5,\gamma^\mu\}=0$ to simplify the algebra, and also changed to the dual Feynman variables 
(see Eq.~(B.8) in Ref.~\cite{Scimemi:2019gge}).

The expression in the parenthesis in \eqref{app:A:2} gives rise to a genuine twist-three contribution. 
It can be rewritten in terms of quark-antiquark-gluon operators using QCD equations of motion.
For example, 
\begin{align}
q(\bar \alpha z)\overleftarrow{\slashed{\partial}}\slashed{z}\gamma^\mu_T \gamma^5q(\beta z)
&=
q(\bar \alpha z)\(\overleftarrow{\slashed{D}}-ig \slashed{A}(\bar \alpha z)\)\slashed{z}\gamma^\mu_T \gamma^5q(\beta z)
\notag\\&=
ig \int_{-\infty}^{\bar \alpha}\!\!d\sigma\,  q(\bar \alpha z)  F_{\nu z}(\sigma z)\gamma^\nu\slashed{z}\gamma^\mu_T \gamma^5q(\beta z)
\notag\\&=
\int_{-\infty}^{\bar \alpha}\!\!d\sigma\, \mathbb{T}_+^\mu(\bar \alpha z,\sigma z,\beta z)+\mathcal{O}(z^2),
\end{align}
where $\mathbb{T}$ is defined in Eq.~(\ref{def:qGq-operators+}).

The remaining contributions in Eq.~\eqref{app:A:0} are computed in the same way. We obtain
\begin{eqnarray}
D_{(a)}&=&(1-\epsilon)\mathbb{N}C_F\int_0^1d\alpha \biggl\{2\alpha \mathcal{O}^{\gamma_T^\mu \gamma^5}(\bar \alpha z,0)
\\\nn && 
+
\alpha \bar \alpha \biggl[\int_{-\infty}^\alpha d\beta\, \mathbb{T}_-(z,\beta z, \alpha z)+
\int_{-\infty}^1 d\beta\, \mathbb{T}_+(z,\beta z, \alpha z)\biggr]\biggr\},
\\
D_{(b)}&=&-(1-\epsilon)\frac{N_c}{2}\mathbb{N}\int_0^1 d\alpha \biggl\{ \int_{-\infty}^\alpha {\!d\beta\,}\alpha \bar \alpha\Big(
\mathbb{T}_+(z,\beta z,\bar \alpha z)+\mathbb{T}_-(z,\beta z, \alpha z)\Big)
\\\nn &&+
\int_{\alpha}^1 {\!d\beta\,}\alpha \bar \beta\Big(
\mathbb{T}_+(z,\beta z,\bar \alpha z)+\mathbb{T}_-(z,\beta z, \alpha z)\Big)\biggr\},
\\ 
D_{(c)}\!+\!D_{(d)}&=&\frac{\mathbb{N}}{2N_c}\int_0^1 d\alpha \biggl\{-(1-\epsilon) 
\int_{-\infty}^\alpha \!\!d\beta\, \alpha\bar \alpha
\Big(\mathbb{T}_+(z,\beta z, \alpha z)+\mathbb{T}_-(z,\beta z, \alpha z)\Big)
\\\nn && +
\int_{0}^\alpha \!d\beta\, \frac{\beta}{2}\Big[
(2-\beta+\beta \epsilon)\mathbb{T}_+(z,\beta z,\alpha z)-(\beta+\epsilon(2-\beta))\mathbb{T}_-(z,\beta z, \alpha z)\Big]
\\\nn && +
\int_{0}^\alpha \!d\beta\, \frac{\beta}{2}\Big[
(\beta+\epsilon(2-\beta))\mathbb{T}_+(\bar \alpha z,\bar \beta z,0)-(2-\beta+\epsilon \beta)\mathbb{T}_-(\bar \alpha z,\bar \beta z,0)\Big]
\\\nn && +
\int_\alpha^1 \!d\beta\, \frac{\alpha}{2}\Big[
(\alpha+(2-\alpha)\epsilon)\mathbb{T}_+(z,\beta z,\bar \alpha z)-(2-\alpha+\epsilon\alpha)\mathbb{T}_-(z,\beta z,\bar \alpha z)\Big]\biggr\},
\end{eqnarray}
where we have made a total shift of the operator position and performed a series of changes of variables to present the expression in a simpler form. 
Summing up everything, we observe that the gauge-dependence due to the choice of the ``retarded'' integration limit $-\infty$ cancels, and 
the final result for this set of diagrams reads
\begin{align}\label{app:A:final}
D&=(1-\epsilon)\mathbb{N}C_F\int_0^1d\alpha \, 2\alpha \,\mathcal{O}^{\gamma_T^\mu \gamma^5}(\bar \alpha z,0)
\notag\\&\quad 
+\mathbb{N}\int_0^1 d\alpha \biggl\{ C_F
\int_{\alpha}^1 d\beta\, \frac{\alpha}{2}(2-\alpha+\alpha\epsilon)
\Big(\mathbb{T}_+(z,\beta z, \alpha z)-\mathbb{T}_-(z,\beta z, \alpha z)\Big)
\notag\\ &\quad
 -\frac{1}{2N_c}\int_{0}^\alpha d\beta \frac{\beta}{2}\Big[
(2-\beta+\beta \epsilon)\mathbb{T}_+(z,\beta z,\alpha z)-(\beta+\epsilon(2-\beta))\mathbb{T}_-(z,\beta z, \alpha z)\Big]
\notag\\ &\quad
 -\frac{1}{2N_c}
\int_{0}^\alpha d\beta\, \frac{\beta}{2}\Big[
(\beta+\epsilon(2-\beta))\mathbb{T}_+(\bar \alpha z,\bar \beta z,0)-(2-\beta+\epsilon \beta)\mathbb{T}_-(\bar \alpha z,\bar \beta z,0)\Big]
\notag\\ &\quad
 +\frac{N_c}{2} \int_\alpha^1 d\beta\, \frac{\alpha}{2}\Big[
\big((2\beta-\alpha)(1-\epsilon)-2\big)\mathbb{T}_+(z,\beta z,\bar \alpha z)
\notag\\&\qquad\qquad\qquad\qquad+\big((2\beta-\alpha)(1-\epsilon)+2\epsilon\big)\mathbb{T}_-(z,\beta z,\bar \alpha z)\Big]\biggr\}.
\end{align}
This expression coincides with Eq.~(\ref{app:B:final}) after appropriate change of variables. We have checked that the same result 
is obtained using the advanced axial gauge. Altogether, these different versions of  the calculation provide an excessive check of the result.

\subsection{Method B}
In this appendix we explain the calculation of the same contribution, with a gluon exchange between the quarks, 
using the technique of Ref.~\cite{Balitsky:1987bk}. In this case we have only one Feynman diagram
\begin{align}
D\,=\,
\begin{minipage}{3cm}
\includegraphics[width=3.0cm]{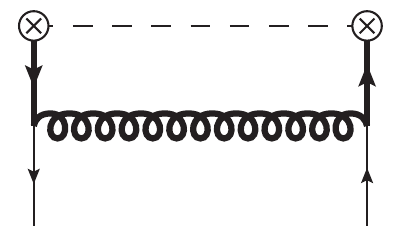}
\end{minipage}
&=
(ig)^2\int d^dy\int d^dx \, \bar q(x) 
\contraction{}{\slashed{A}}{(x) q(x)\bar q(z) \gamma_T^\mu \gamma_5 q(0) \bar q(y)}{\slashed{A}}
\contraction{\slashed{A}(x)}{q}{(x)}{\bar q}
\contraction{\slashed{A}(x) q(x)\bar q(z) \gamma_T^\mu \gamma_5 }{q}{(0) }{\bar q}
\slashed{A}(x) q(x)\bar q(z) \gamma_T^\mu \gamma_5 q(0) \bar q(y)  \slashed{A}(y)q(y)
\end{align}
where the Wick contractions correspond to the quark and gluon propagators in the background field~\cite{Balitsky:1987bk}
\begin{align}
\label{q-prop}
\contraction{}{q^a}{(x)}{q}
q^a(x)\bar q^b(y)
&=
\frac{i}{2\pi^\dhalf}
\frac{\slashed{\Delta} \Gamma(\dhalf)}{[-\Delta^2]^\dhalf}[x,y]^{ab}
\notag\\&\quad
+ \frac{ig}{16\pi^\dhalf} \frac{\Gamma(\dhalf-1)}{[-\Delta^2]^{\dhalf-1}}
\int_0^1\!du\, \Big[ \bar u \slashed{\Delta} \sigma F(ux + \bar u y) + u \sigma F(ux+\bar u y) \slashed{\Delta}\Big]+\ldots
\end{align}
where  $\Delta = x-y$,  $\sigma F = \sigma^{\alpha\beta}F^A_{\alpha\beta}t_{ab}^A$, and
\begin{align}
\label{g-prop}
\contraction[6pt]{\,}{A}{~\,(x)}{A}
A^A_{\alpha}(x)A^B_{\beta}(y)
&= 
 - \frac{1}{4\pi^\dhalf} g_{\alpha\beta} 
\frac{\Gamma(\dhalf-1)}{[-\Delta^2]^{\dhalf-1}}[x,y]^{AB}
- \frac{i}{8\pi^\dhalf}
 \frac{\Gamma(\dhalf-2)}{[-\Delta^2]^{\dhalf-2}}
\int_0^1\!du\, gF^{AB}_{\alpha\beta}(ux+ \bar u y) + \ldots
\end{align}
with $F^{AB}_{\mu\nu} = i f^{AEB} F^E_{\mu\nu}$. In both expressions, 
the ellipses stand for terms $\mathcal{O}([-\Delta^2]^{3-\dhalf})$ which do not contribute to 
our accuracy.%
\footnote{Strictly speaking, the gluon fields in \eqref{q-prop}, \eqref{g-prop} have to be decorated with 
the Wilson lines,  
$F(ux+\bar u y)\mapsto [x,ux+\bar u y]F(ux+\bar u y)[ux+\bar u y]$. They can be dropped, however, to our accuracy.} 
Note that the Wilson line in \eqref{g-prop} is taken in the adjoint representation.
For clarity we use capital letters $A,B,\ldots$ for color octet indices.

For bookkeeping purposes it is convenient to split the calculation in the contributions with and without the background gluon field
in the propagators, similarly to Fig.~\ref{fig:diagrams}.
The contribution (a) is by far the most complicated one; let us consider it in some detail.
\begin{align}\label{app:b:1}
 D_{(a)} &= \frac{ g^2}{16\pi^{\frac{3d}{2}}} \int\! d^dy\, \int\! d^dx\, \frac{\Gamma(\dhalf-1)}{[-(x-y)^2]^{\dhalf-1}}
\frac{\Gamma(\dhalf)}{[-(x-z)^2]^{\dhalf}}  \frac{\Gamma(\dhalf)}{[-y^2]^{\dhalf}}
\notag\\&\qquad\times
\bar q(x) \gamma^\sigma (\slashed{x}-\slashed{z})[x,z]\gamma_T^\mu\gamma_5 
 \slashed{y}[0,y]\gamma_\sigma  t^A [x,y]_{AB} t^B q(y) 
\notag\\&=
\frac{ g^2}{16\pi^{\frac{3d}{2}}} 
\int\! d^dy\, \int\! d^dx\, \int_0^1\!du\, u^{\dhalf-2}\bar u^{\dhalf-1} \frac{\Gamma(d-1)}{[-y^2-u\bar u x^2]^{d-1}} 
\frac{\Gamma(\dhalf)}{[-(x-z)^2]^{\dhalf}}
\notag\\&\qquad\times
\bar q(x) \gamma^\sigma (\slashed{x}-\slashed{z})[x,z]\gamma_T^\mu\gamma_5 
(\slashed{y}+u \slashed{x})[0,y+ux] \gamma_\sigma t^A [x,y+ux]_{AB} t^B  q(y+ux) 
\end{align}
The basic idea is that the remaining (classical) fields in the integral can be expanded as
\begin{align}
    q(y+ux) = q(ux) + y^\xi \partial_\xi q(ux) + \ldots
\end{align}
after which the $x$-integration becomes trivial, 
and it is easy to convince oneself that the terms with more than one derivative produces 
corrections $\sim \mathcal{O}(z^2)$ and can be dropped. Wilson lines also have to be expanded and to this end
using Fock-Schwinger gauge for the classical gluon field $x_\xi A^\xi(x)=0$, $A^\xi(0)=0$, proves to be
very convenient. In this gauge $[0,y+ux] = \II$ and
\begin{align}
\label{FSfield}
       A_\xi(x) = \int_0^1d\alpha\,\alpha\, x^\rho F_{\rho \xi} (\alpha x)  
\end{align} 
so that
\begin{align}
\label{glueWL}
   t^A [x,y+ux]_{AB} t^B  &= C_F + ig\frac{N_c}{2} \int_0^1\!dt \,(\bar u x -y)^\xi A_\xi(\bar t (y+ux) + t x)  +\ldots
\notag\\&=
 C_F + ig\frac{N_c}{2} \int_0^1\!dt \, \int_0^1\!d\alpha\,\alpha\, x^\xi y^\eta F_{\eta\xi}(\alpha [\bar t (y+ux) + t x]) +\ldots
\end{align}
Taking the $x$-integral, we repeat the same procedure to do the $y$-integration: 
combine the remaining two propagators introducing another Feynman parameter, shift the integration variable, and expand the 
classical fields along the $z$-direction (the antiquark field and the $[x,z]$ Wilson line). In this way one ends up with four terms:
\begin{enumerate}
\item A quark-antiquark contribution without derivatives on the fields,
\item A quark-antiquark contribution with one derivative, 
$\bar q(vz)\stackrel{\leftarrow}{\partial} q(uz)$ or $\bar q(vz)\stackrel{\rightarrow}{\partial} q(uz)$ 
\item A gluon from the expansion of the Wilson line in the gluon propagator, \eqref{glueWL},
\item A gluon from the expansion of the Wilson line in the quark propagator from $z$ to $x$.
\end{enumerate}   
The last term can be handled by rewriting $\partial_\xi = D_\xi + i g A_\xi$, using \eqref{FSfield} and 
the operator identity 
\begin{align}
\partial_\xi \bar q(vz) [vz,uz]\slashed{z}\Gamma q(uz) &= 
 \bar q(vz)\biggl\{\stackrel{\leftarrow}{D_\xi} - i \int_{u}^vdt\, g F_{z \xi}(tx) + \stackrel{\rightarrow}{D_\xi}  
\biggr\}\Gamma q(uz) 
\end{align}
so that neglecting total derivatives
\begin{align}
 \bar q(vz ) \Big[\slashed{z} \gamma_T^\mu\stackrel{\leftarrow}{\slashed{D}} \Big]\gamma_5 q(uz)  
&=  i  \int_{u}^vdt\,\bar q(vz)\,\slashed{z}\gamma_T^\mu gF_{z\xi}(tz)\gamma^\xi \gamma_5q(uz)\,, 
\notag\\
\bar q(vz ) \Big[ \stackrel{\rightarrow}{\slashed{D}} \gamma_T^\mu  \slashed{z}  \Big]\gamma_5 q(uz) 
&=  i  \int_{u}^vdt\,\bar q(vz)\, g F_{z\xi}(tz)\gamma^\xi\gamma_T^\mu\slashed{z}\gamma_5  q(uz)\,. 
\end{align}
Here $F_{z\xi} = z^\eta F_{\eta\xi}$.
In this way we obtain ($d=4-2\epsilon$)
\begin{align}
 D_{(a1)} &= 2 C_F \mathbb{N} (1-\epsilon) \int_0^1dv\!\int_0^v\!du\,\bar q(vz ) \gamma_T^\mu \gamma_5 q(uz)\,, 
\notag\\
D_{(a2)} &=
ig C_F \mathbb{N}
\int_0^1\!dv \!\int_0^v du\!
\int_{u}^v\!dt\,\bar q(vz)
\Big\{\big[\bar u  + \epsilon u \big] \gamma^\rho\gamma_T^{\mu}
- \big[v + \epsilon\bar v\big] \gamma_T^{\mu}\gamma^\rho 
\Big\} F_{z\rho}(tz)\slashed{z}\gamma_5 q(uz) 
\notag\\&\quad
- ig C_F \mathbb{N}
\int_0^1\!dv \!\int_0^v\! du\!\int_u^v\!dt\,\frac{t}{v}\,
\bar q(vz ) 
\Big\{\big[\bar v + \epsilon v\big] \gamma^\rho \gamma_T^{\mu} 
- \big[v+ \epsilon\bar v\big]\gamma_T^{\mu} \gamma_\rho \Big\}F_{z\rho}(tz)\slashed{z} \gamma_5q(uz) 
\notag\\&\quad
- ig C_F \mathbb{N}
\int_0^1\!dv \!\int_0^v\! du \left(\frac{1}{v}-\frac{1}{u}\right) \int_0^u\!dt\,t\,
\bar q(vz ) 
\Big\{\gamma^\rho \gamma_T^{\mu} 
- \epsilon \gamma_T^{\mu} \gamma_\rho \Big\} F_{z\rho}(tz)\slashed{z} \gamma_5q(uz)\,, 
\notag\\
  D_{(a3)} &=
- ig \frac{N_c}{2} \mathbb{N}
\int_0^1\!dv\, \int_0^v\!du\,  
 \int_u^v dt \, \left(1-\frac{t}{v}\right)
\bar q(vz) \big[\gamma^\rho  \gamma_T^\mu  - \epsilon \gamma_T^\mu  \gamma^\rho\big]
F_{z \rho }(t z  ) \slashed{z} \gamma_5q(uz) 
\notag\\&\quad
- ig \frac{N_c}{2} \mathbb{N}
\int_0^1\!dv\, \int_0^v\!du\, 
\int_0^u dt \, \left(\frac{t}{u}- \frac{t}{v}\right) 
\bar q(vz)\big[\gamma^\rho  \gamma_T^\mu  - \epsilon \gamma_T^\mu  \gamma^\rho\big]
F_{z \rho }(t z  ) \slashed{z} \gamma_5q(uz)\,, 
\notag\\
 D_{(a4)} &=
ig  \frac{1}{2N_c} \mathbb{N}
\int_0^1\!dv\! \int_0^v \!du\! \int_0^v \! dt\, t\,  
\bar q(vz) \big[\gamma_T^\mu \gamma^\rho - \epsilon \gamma^\rho \gamma_T^\mu \big]
F _{z \rho}\big(t z \big)  \slashed{z}\gamma_5 q(uz) 
\notag\\&\quad -
ig  \frac{1}{2N_c} \mathbb{N}
\int_0^1\!dv \frac{\bar v}{v} \int_0^v\!du\,
\int_0^v \! dt\, t\,  
\bar q(vz) \big[ \gamma^\rho  \gamma_T^\mu  - \epsilon \gamma_T^\mu  \gamma^\rho\big]
F _{z \rho}\big(t z \big) \slashed{z}\gamma_5 q(uz) 
\notag\\&\quad +
ig  \frac{1}{2N_c} \mathbb{N}
\int_0^1\!dv\,\frac{v}{\bar v }  \int_0^v \!du\! 
\int_v^1\! dt\,\bar t\, 
\bar q(vz) \big[\gamma_T^\mu \gamma^\rho - \epsilon \gamma^\rho \gamma_T^\mu \big]
F _{z \rho}\big(t z \big)  \slashed{z}\gamma_5 q(uz) 
\notag\\&\quad -
ig  \frac{1}{2N_c} \mathbb{N}
\int_0^1\!dv \int_0^v\!du\, 
\int_v^1\! dt\, \bar t\,
\bar q(vz) \big[ \gamma^\rho  \gamma_T^\mu  - \epsilon \gamma_T^\mu  \gamma^\rho\big]
F _{z \rho}\big(t z \big) \slashed{z} \gamma_5 q(uz)\,, 
\end{align}
where $\mathbb{N}$ is defined in Eq.~(\ref{app:A:N}).

The remaining contributions (b)--(d) are much simpler; their calculation is straightforward. We obtain
\begin{align}
  D_{(b)} &= 0\,,
\notag\\
  D_{(c)} &= 
- ig \frac{1}{2N_c} \mathbb{N}
\int_0^1\!dv\! \int_0^v\!du\! \int_0^u\! dt\,\left(1-\frac{t}{u}\right)
\bar q(vz)
 \Big[\gamma^\rho \gamma_T^\mu  - \epsilon  \gamma_T^\mu  \gamma^\rho \Big] F_{z\rho}(tz)  \slashed{z}\gamma_5q(uz) \, ,
\notag\\
 D_{(d)} &=
ig \frac{1}{2N_c} \mathbb{N}
\int_0^1\!dv\,\int_0^v\!du\,\int_v^1 dt  \left(1-\frac{\bar t}{\bar v}\right)
\bar q(vz)
\Big[\gamma_T^\mu  \gamma^\rho  - \epsilon \gamma^\rho \gamma_T^\mu \Big] F_{z\rho}( t z)  \slashed{z}\gamma_5 
q(uz) \, .
\end{align}
Finally, summing up everything, we get
\begin{align}\label{app:B:final}
D &= 
2 C_F \mathbb{N}\,(1-\epsilon)
\int_0^1dv\!\int_0^v\!du\,\bar q(vx ) \gamma_T^\mu \gamma_5 q(ux)
\notag\\& \quad
+
ig \left(C_F\!-\!\tfrac12 N_c\right) \mathbb{N}
\int_0^1\!\!dv\!\! \int_0^v\!\!du\!\!
\int_0^u \!\! dt\,  
\bar q(vx) \Big\{[\bar t + \epsilon t] \gamma^\rho  \gamma_T^\mu  - [t+ \epsilon\bar t] \gamma_T^\mu \gamma^\rho\Big\}
F _{x \rho}\big(t x \big)\slashed{x}\gamma_5q(ux) 
\notag\\& \quad
+
ig \left(C_F\!-\!\tfrac12 N_c\right) \mathbb{N}
\int_0^1\!\!dv\!\! \int_0^v\!du\!\! \int_v^1\!\! dt\,
\bar q(vx)
\Big\{ [\bar t + \epsilon t] \gamma^\rho  \gamma_T^\mu  - [t + \epsilon\bar t] \gamma_T^\mu \gamma^\rho \Big\}
F _{x \rho}\big(t x \big)\slashed{x}\gamma_5q(ux) 
\notag\\& \quad
+ ig C_F  \mathbb{N}
\int_0^1\!dv \!\int_0^v du\!
\int_{u}^vdt\,\bar q(vx)\,
\Big\{\big[ \bar u  + \epsilon u \big] \gamma^\rho\gamma_T^\mu 
 - \big[v + \epsilon\bar v\big]\gamma_T^\mu \gamma^\rho 
\Big\}F_{x\rho}(tx)\slashed{x}\gamma_5q(ux) 
\notag\\& \quad 
-ig \frac{N_c}{2} \mathbb{N}
\int_0^1\!dv\, \int_0^v\!du\,  
 \int_u^v dt \, 
\bar q(vx)\Big\{[ \bar t + \epsilon t]\gamma^\rho \gamma_T^\mu  - [t + \epsilon\bar t ] \gamma_T^\mu \gamma^\rho\Big\}
F _{x \rho}\big(t x \big)\slashed{x}\gamma_5 q(ux)\, ,
\end{align}
where in the first line one still needs to separate twist-two and twist-three contributions 
\begin{align}
\bar q(vz ) \gamma_T^\mu \gamma_5 q(uz) & = [\bar q(vz ) \gamma_T^\mu \gamma_5 q(uz)]_{\mathrm{tw2}} + [\bar q(vz ) \gamma_T^\mu \gamma_5 q(uz)]_{\mathrm{tw3}} 
\end{align}
as shown in Eqs.\,\eqref{t=2}, \eqref{t=3}. The result in \eqref{app:B:final} coincides with Eq.~\eqref{app:A:final} after the appropriate change of variables.


\bibliography{bibFILE}

\bibliographystyle{JHEP}


\end{document}